\documentclass[aip,jcp,reprint,amsmath,amssymb,floatfix,citeautoscript,nofootinbib]{revtex4-2}
\usepackage{cancel}
\usepackage{amsmath}
\usepackage{physics}
\usepackage{graphicx}
\usepackage{amssymb}
\usepackage{amsthm}
\usepackage{bm}
\usepackage{dcolumn}
\usepackage{multirow}
\newcolumntype{x}[1]{D{.}{.}{#1}}

\usepackage{ragged2e}
\usepackage{newtxtext,newtxmath}
\allowdisplaybreaks

\usepackage{color}
\usepackage[usenames,dvipsnames]{xcolor}
\definecolor{myblue}{rgb}{0,0,1}
\usepackage[breaklinks=true,colorlinks=true,linkcolor=myblue,urlcolor=myblue,citecolor=myblue]{hyperref}

\let\vr\undefined
\newcommand{\vr}{{\bm{r}}}
\newcommand{\vR}{{\bm{R}}}
\newcommand{\vq}{{\bm{q}}}
\newcommand{\eone}{\mathcal{E}_{\mathrm{D}^0\mathrm{A}^0}}
\newcommand{\etwo}{\mathcal{E}_{\mathrm{D}^+\mathrm{A}^-}}

\usepackage{tabularx}

\begin{document}

\title{
Diabatic states of charge transfer with constrained charge equilibration 
}

\author{Sohang Kundu}
\email{sohangkundu@gmail.com}
\affiliation{Department of Chemistry, Columbia University, New York, New York 10027, USA}
\author{Hong-Zhou Ye}
\affiliation{Department of Chemistry, Columbia University, New York, New York 10027, USA}
\author{Timothy C. Berkelbach}
\email{t.berkelbach@columbia.edu}
\affiliation{Department of Chemistry, Columbia University, New York, New York 10027, USA}
\affiliation{Initiative for Computational Catalysis, Flatiron Institute, New York, New York 10010, USA}

\begin{abstract}
Charge transfer (CT) processes that are electronically non-adiabatic are ubiquitous in chemistry, biology, and materials science,
but their theoretical description requires diabatic states or adiabatic excited states. For complex systems, these latter states 
are more difficult to calculate than the adiabatic ground state.
Here, we propose a simple method to obtain diabatic states, including energies and charges, by constraining the atomic charges within
the charge equilibration framework.
For two-state systems, the exact diabatic coupling can be determined, from which the adiabatic excited-state energy can also be calculated.
The method can be viewed as an affordable alternative to constrained density functional theory (CDFT), and so we call it constrained charge equilibration (CQEq).
We test the CQEq method on the anthracene-tetracyanoethylene CT complex and the reductive decomposition of ethylene carbonate on a lithium metal surface.
We find that CQEq predicts diabatic energies, charges, and adiabatic excitation energies in good agreement with CDFT, and we propose that CQEq is promising for combination with machine learning force fields to study non-adiabatic CT in the condensed phase.
\end{abstract}
 
\maketitle

\section{Introduction}
The charge equilibration (QEq) method~\cite{Rappe1991}, which is based on the
concept of electronegativity equalization~\cite{Sanderson1976,Mortier1986},
provides a simple and affordable way to predict atomic charges and
electrostatic energies of molecular systems.  Force fields including QEq
charges and energies have been developed for over two
decades~\cite{Rick1996,Patel2004,Patel2006,Jing2019,Caldeweyher2019} and many
of these FFs are reactive~\cite{Duin2001,Senftle2016,Li2021}.  In recent years,
the introduction of QEq in machine learning FFs (MLFFs)
\cite{Behler2021,Unke2021} has enabled charge-aware molecular dynamics (MD)
simulations at near \textit{ab initio} accuracy, but such
MLFFs\cite{Ko2021,Deng2023,Kovacs2023} are usually designed to describe only
the Born-Oppenheimer ground state.  Although some charge transfer (CT)
reactions happen adiabatically on the ground state, in this work we are
interested in non-adiabatic CT processes.

One approach to MLFFs for non-adiabatic dynamics is to train separate models
that predict ground and excited state energies, and possibly non-adiabatic
couplings~\cite{Chen2018,Dral2018,Westermayr2020,Dupuy2024}.
Such an approach requires additional training data from excited-state electronic
structure theory, and training a model to predict non-adiabatic couplings can be
challenging~\cite{Richardson2023}.
Here, we propose and test a much simpler approach.
Specifically, we obtain diabatic states, including energies and charges,
by minimizing the QEq electrostatic energy with constraints placed on the
atomic charges of donor and acceptor molecules.
This method can be considered an approximation to constrained density functional
theory (CDFT)~\cite{Kaduk2011}, and so we call it constrained charge equilibration (CQEq).
In principle, the QEq parameters could be reoptimized to provide accurate results
for diabatic states. Remarkably, we find that QEq parameterizations that are designed for
adiabatic ground states perform quite well in predictions for diabatic states. 

In Sec.~\ref{sec:theory},
we present the theory behind CQEq. In Sec.~\ref{sec:results}, we apply CQEq to
an organic donor-acceptor complex and the reductive decomposition of an adsorbate
on a metal surface. In both cases, we benchmark the CQEq diabatic and
adiabatic energies against those obtained from CDFT, and for the latter example
we also compare the diabatic state charge distributions. Concluding
remarks are presented in Sec.~\ref{sec:conc}.

\section{Theory}
\label{sec:theory}

In QEq methods, the ground-state potential energy depends on
the nuclear positions $\vR = (\vR_1,\ldots,\vR_N)$ 
and the atomic charges $\vq = (q_1,\ldots,q_N)$. The energy is 
decomposed as
\begin{equation}
U_0(\vR, \vq) = U_\mathrm{short}(\vR) + U_\mathrm{elec}(\vR,\vq).
\end{equation}
The electrostatic energy is the sum of the energies to create a partial
charge $q_i$ on each atom and a Coulomb energy,
\begin{equation}
\label{eq:qeq}
\begin{split}
U_\mathrm{elec}(\vR,\vq)
    &= \sum_i \left[\chi_i q_i 
        + \frac{1}{2} J_{ii} q_i^2\right] \\
    &\hspace{1em} + \frac{1}{2} \sum_{i\neq j} \iint d\vr_1 d\vr_2 
        \frac{\rho_i(\vr_1) \rho_j(\vr_2)}{r_{12}}
\end{split}
\end{equation}
where $\rho_i(\vr)$ is an atomic charge density centered on atom $i$ integrating
to charge $q_i$.

The atomic charge distributions are commonly modeled with spherically symmetric
Slater-type orbitals~\cite{Rappe1991} (STOs)
or Gaussians~\cite{Caldeweyher2019} with atom-specific exponents $\zeta_i$.
For example, with STOs, which we use in this work, the atomic charge density is
$\rho_i(\vr) = q_i|\phi_{i}(\vr)|^2$, with
$\phi_{i}(\vr) = N_{i} r^{n_i-1} e^{-\zeta_i r}$,
where $n_i$ is the principal quantum number of the valence electrons in atom $i$,
and the Coulomb integrals are calculated numerically.
Only three parameters per atom are needed: $\chi_i, J_{ii}, \zeta_i$,
which are the electronegativity, hardness, and exponent. 
Finally, the charges $q_i$ are obtained by minimizing the electrostatic energy
$U_\mathrm{elec}$ subject to the constraint $\sum_i q_i = Q$, where $Q$ is the
total charge of the system. 

We now ask whether the ground-state QEq energy function is accurate away
from its minimum (a priori, there is no reason it should be). Specifically, we propose
the CQEq method, and show how it can be used to constract diabatic
states, including diabatic charge distributions, energies, and couplings---
and, from these, adiabatic excited states, and non-adiabatic couplings.
Analogous to CDFT, diabatic states are accessed by placing a constraint on the 
fluctuating charges.
For example, consider an electron donor-acceptor system with the two diabatic states D$^0$A$^0$
and D$^{+}$A$^{-}$. We calculate the energy and charge distribution of these diabatic states
by minimizing the QEq electrostatic energy
subject to the addition constraint $\sum_{j\in \mathrm{D}} q_j=Q_\mathrm{D}$,
where $Q_\mathrm{D}$ is the target charge on the donor (0 or 1).
These diabatic energies and charge distributions would be useful in studies of 
electron transfer in polar solvents,
especially in the diabatic, Marcus-like limit~\cite{Kuharski1988,Sit2006}.

Furthermore, from the diabatic energies $\eone$,
$\etwo$
and the adiabatic ground-state energy $U_0$,
we can infer the magnitude of the diabatic coupling,
\begin{equation}
V(\vR) 
    = \sqrt{\left[U_0(\vR) - \eone(\vR)\right]
            \left[U_0(\vR) - \etwo(\vR)\right]},
\end{equation}
and the adiabatic excited-state energy as the higher-energy eigenvalue of the diabatic
Hamiltonian matrix,
\begin{equation}
\begin{split}
U_1(\vR) &= \frac{1}{2}\Big[ \eone(\vR) + \etwo(\vR) \\
    &\hspace{2em} + \sqrt{[\eone(\vR)-\etwo(\vR)]^2 + 4 [V(\vR)]^2}\Big]\\
    &= \eone(\vR) +  \etwo(\vR) -  U_{0}(\vR),
\end{split}
\end{equation}
where the last equality follows from the basis invariance of the matrix trace. 
The adiabatic CT excitation energy is defined as the vertical energy gap between
adiabatic states, $\Delta U(\vR) = U_1(\vR) - U_0(\vR)$. 
Note that we assume that the non-electrostatic short-range part of the potential energy $U_\mathrm{short}(\vR)$
is identical in the ground and excited states.
Finally, an approximate non-adiabatic coupling vector can be calculated through the derivatives of the
adiabatic eigenvector coefficients. Although we will not perform numerical tests of the non-adiabatic
coupling vector, we return to this point in Sec.~\ref{sec:conc}.

The CQEq method is closely related to the empirical valence bond (EVB) method~\cite{Warshel1980,Kamerlin2011}
or the multiconfiguration molecular mechanics method~\cite{Kim2000},
but it serves a slightly different purpose. In EVB methods, non-reactive force fields associated with
fixed charge distributions are commonly
used for the diabatic energies; diagonalization of a Hamiltonian with parameterized diabatic couplings
then yields a ground state corresponding to a reactive force field with an adiabatic charge distribution.
In CQEq, we assume the existence of a reactive, ground-state force field with an adiabatic charge distribution, 
such as those available from recent MLFFs \cite{Behler2021} using QEq charges. We then use CQEq to calculate the 
diabatic ``force field'' energies and, from the diabatic and adiabatic energies, the diabatic couplings.
In this sense, our work is most similar to the EVB study in Ref.~\onlinecite{Hong2006}, wherein CDFT was used to calculated the
diabatic states and DFT was used to calculate the adiabatic ground state; the inferred diabatic couplings
were then used to test the common assumption of the EVB approach that the couplings are not strongly environment-dependent
and are thus transferable~\cite{Warshel1980}.

Before presenting our results, we note that the QEq method, upon which CQEq is built, 
is well-known to predict unphysical, long-range charge transfer,
even for systems that should dissociate into closed shell species.
This pathology can be alleviated using environmentally dependent electronegativities,
$\chi_i(\vR)$, as commonly done in MLFFs~\cite{Ghasemi2015,Ko2021,Jacobson2022}.
To affordably mimic accurately trained environmentally dependent electronegativities,
we generate ``exact'' values via 
\begin{equation}
    \label{eq:chiR}
    \chi_i(\vR) = - \sum_j J_{ij} \tilde{q}_j
\end{equation}
where $\tilde{q}_j$ are atomic partial charges obtained from a reference
adiabatic ground-state quantum chemistry calculation.   
By construction, the ``exact'' charges $\tilde{q}_i$ minimize 
$U_\mathrm{elec}$. As long as the reference method correctly describes
long-range charge transfer, Eq.~\ref{eq:chiR} will necessarily eliminate the
unphysical behavior of QEq. 
For all results presented in the next section, we use previously reported values~\cite{Rappe1991,Caldeweyher2019}
for the hardness $J_{ii}$ and exponents $\zeta_{i}$, which are presented in the SI.

\begin{figure}[t]
    \includegraphics[scale=0.85]{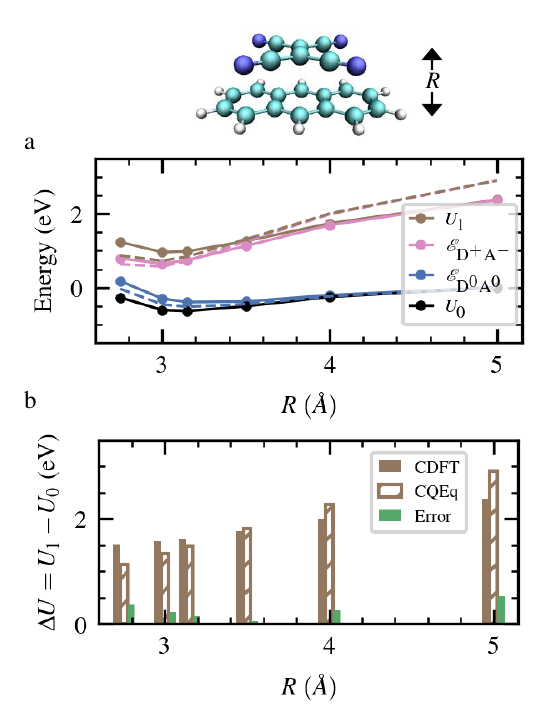}
    \caption[\centering]{Diabatic and adiabatic charge-transfer energies for
the organic donor-acceptor complex of anthracene and TCNE. In panel (a), solid
lines show CDFT energies and dashed lines show CQEq energies.  In panel (b),
adiabatic CT excitation energies calculated using CQEq and CDFT, as well as the
absolute difference between the two, are shown.
} 
    \label{fig:TA}
\end{figure}

\begin{figure}[t]
    \includegraphics[scale=0.9]{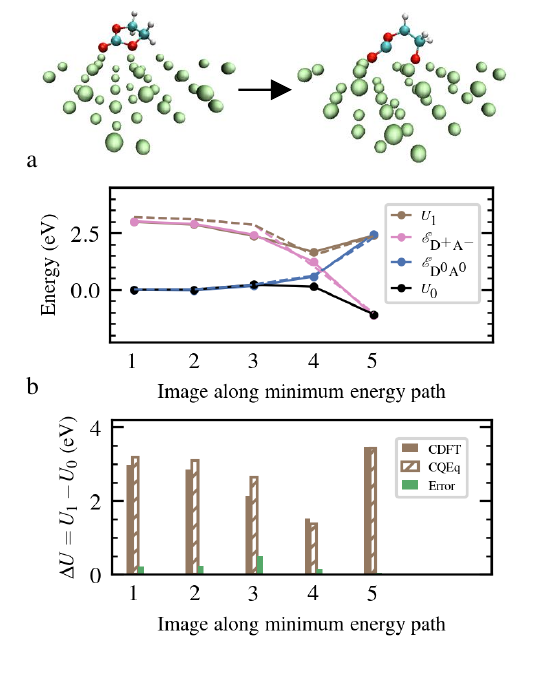}
    \caption[\centering]{Same as Figure \ref{fig:TA} but for the decomposition of 
    ethylene carbonate (EC) on the surface of lithium metal. Molecular clusters
containing 40 Li atoms and EC were constructed from periodic images along the
minimum energy pathway.
} 
    \label{fig:ECLi}
\end{figure}

\section{Results}
\label{sec:results}

Some of the simplest illustrations of CT with diabatic and
adiabatic states are those arising in diatomic molecules~\cite{VanVoorhis2010}. However, since
diatomics usually dissociate into open shell fragments, most DFT functionals 
predict spurious long-range, fractional CT due to their absence of a derivative
discontinuity~\cite{Perdew1982,MoriSanchez2014} (we emphasize that this problem in DFT
is distinct from the long-range CT pathology of QEq). 
In this case, one can still use a
higher level reference method to extract exact ground-state QEq parameters. 
We have performed CQEq calculations using CASSCF reference charges 
for LiF, LiCl, and NaCl, and the results are provided in the Supporting Information. 
Although the simplicity of these diatomics allows us to obtain some insights
into the exact QEq parameters needed to reproduce the CASSCF adiabatic excitation
energy, we believe that diatomics are a special case.  

We now use CQEq to calculate diabatic and adiabatic
excitation energies for two more complicated CT systems, both involving polyatomic
donors and acceptors. All reference and benchmarking calculations are performed
using unrestricted DFT and CDFT respectively, with the PySCF~\cite{Sun2017}
quantum chemistry package. The meta-L{\"o}wdin population analysis \cite{Sun2014} is used to
calculate atomic partial charges. Further computational details are summarized
below.

Our first example is the symmetrically stacked organic donor-acceptor complex of anthracene and
tetracyanoethylene (TCNE). Ground-state DFT calculations, performed using the
B3LYP functional~\cite{Becke1988,Lee1988} with D3 dispersion~\cite{Grimme2010,Grimme2011} and the def2-TZVP basis set~\cite{Schaefer1994,Weigend2005}, show that at
equilibrium separation ($R=3.15$~\AA), anthracene transfers $0.27~e$ to TCNE,
in agreement with previous studies \cite{Siddique2020}. The 
neutral and ionic diabats are defined by constraining the charge on anthracene
to be 0 or +1.
Figure~\ref{fig:TA}(a) shows the energies of the adiabatic and
diabatic states as a function of separation $R$, calculated by both CDFT and CQEq. 
Consistent with the relatively small
amount of ground-state CT, we note that the diabatic potential energy surfaces do not cross.
The CQEq results are in good agreement with those from CDFT, differing by less than 0.5~eV
at all separations $R$.
In Fig.~\ref{fig:TA}(b), we compare the adiabatic CT excitation energies $\Delta U = U_1-U_0$.
CQEq correctly predicts the magnitude of the excitation energy to be about 1.5~eV at equilibrium
and to increase with increasing $R$.
The good agreement between CQEq and CDFT demonstrates the accuracy of the 
QEq electrostatic energy expression away from its minimum.

Our second example is the reductive ring-opening decomposition of ethylene carbonate (EC) on the (001) 
surface of lithium metal, which has been previously studied with periodic DFT~\cite{Ebadi2016,Brennan2017,CamachoForero2015}. 
This reaction, which involves electron transfer from metal to molecule, 
is of interest for the formation of the solid-electrolyte interphase in batteries
with a lithium metal anode.
A nudged elastic band calculation with periodic boundary conditions was performed
using Quantum Espresso \cite{Giannozzi2009,Giannozzi2017} with the PBE functional \cite{Perdew1996,Perdew1997}, generating the ground-state minimum energy path,
which has a small barrier of about 0.22~eV.
This minimum energy path was used for subsequent CDFT and CQEq calculations,
for which we approximate the lithium surface with a 40-atom cluster to allow the
use of molecular quantum chemistry software. 
The molecular calculations are performed using the PBE functional and a double-$\zeta$ correlation consistent Gaussian basis set \cite{Ye2022}.

In Fig.~\ref{fig:ECLi}(a), we show the diabatic and adiabatic state energies predicted
by CDFT and CQEq, and in Fig.~\ref{fig:ECLi}(b), we show the adiabatic excitation energies.
The accuracy of CQEq is again quite good, accurately capturing the diabatic state energies over
the entire reaction path and predicting adiabatic excitation energies to an accuracy of better
than 0.5~eV.
Unlike for the anthracene-TCNE complex, the diabatic surfaces now cross as the
ground state changes from neutral to ionic character along the reaction path. 
Interestingly, this changing character appears responsible for the ground-state reaction energy
barrier.

\begin{figure}[t]
    \includegraphics[scale=0.95]{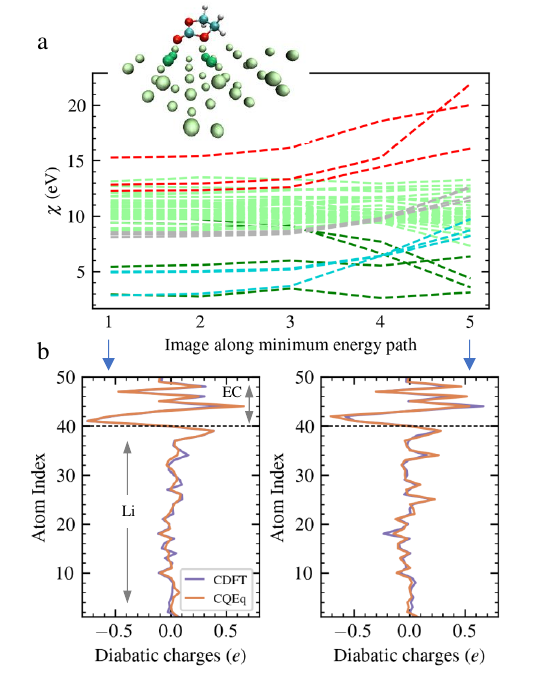}
    \caption[\centering]{(a) Atomic electronegativities calculated using
Eq.~(\ref{eq:chiR}) for the decomposition of ethylene carbonate (EC) on a
40-atom lithium cluster. The color coding of curves follows those in the
molecular graphic: oxygen (red), hydrogen (grey), carbon (cyan), lithiums near
EC (dark green), all other lithiums (light green). (b) Atomic charges in the
diabatic states D$^0$A$^0$ for image 1 (left) and D$^+$A$^-$ for image 5 (left).
Results from CQEq are compared to those of CDFT. 
The dashed black line separates EC atoms from Li atoms, which are arranged in ascending order of their z-coordinates (normal to the Li 001 surface).
} 
    \label{fig:chiqCT}
\end{figure}

Having established the accuracy of CQEq, we conclude with a closer look at the
functioning of the CQEq method and its various components for the latter example
of EC on lithium. 
In Fig.~\ref{fig:chiqCT}(a), we show the electronegativities $\chi_i(\vR)$,
calculated using Eq.~\ref{eq:chiR}, for each image along the reaction pathway. 
Aside from acquiring a geometry dependence, the
electronegativities of atoms of the same type can have vastly different values
due to differences in their local environments. Importantly, the most drastic
variations are observed in the spatial vicinity of the bonds being broken and formed.
For example, the electronegatives of the four lithium atoms closest to the EC
molecule are quite different from the other lithium atoms, and 
the electronegativities of the carbon and oxygen atoms associated with the breaking
bond change suddenly around the fourth image.
These results give some generic insight into the environmental awareness needed
for charge-aware ground-state MLFFs. 
Analogous results for the anthracene-TCNE complex are presented in the Supporting Information.

Finally, in Fig.~\ref{fig:chiqCT}(b), we compare the diabatic charges predicted
by CDFT and CQEq. Specifically, we compare the charges in the neutral diabatic
state of the first image as well as the charges in the ionic diabatic state of the
fifth image.
As evident from Figure \ref{fig:chiqCT}(b), the charge distribution in diabatic states
predicted by CQEq are in excellent agreement with those obtained using CDFT.

\section{Conclusions}
We have introduced an efficient approach to calculate diabatic CT states by
constraining charges within a QEq framework. The central assumption of the CQEq
method is that the QEq electrostatic energy function remains valid (with the
same parameters) away from its global minimum. Our numerical results suggest that
this assumption is very good.

Based on the present results, we propose that the CQEq approach can be used to
study molecular dynamics associated with CT. For example, diabatic free
energies can be straightforwardly calculated by umbrella sampling, and we
expect that the results would be more accurate than those obtained with simple
point-charge force-fields~\cite{Kuharski1988} and more affordable than those
obtained with CDFT~\cite{Sit2006}. For explicit CT dynamics, non-adiabatic molecular dynamics
could be performed in the adiabatic basis, which in principle requires the non-adiabatic coupling vector
or time-derivative coupling.
As mentioned in Sec.~\ref{sec:theory}, these can be obtained with CQEq. 
Alternatively, one could immediately use approximate versions of the non-adiabatic coupling that
depend only adiabatic energies and derivatives~\cite{Baeck2017,Shu2022}, which we think is especially
promising in combination with CQEq.

The CQEq method can be straightforwardly extended to calculate diabatic states associated with more than two
CT configurations, which would be important in complex, condensed-phase systems exhibiting charge diffusion.
However, if one is interested in diabatic couplings or adiabatic excited states, then the CQEq framework is insufficient:
for a system with $n$ electronic states,
the $n$ diabatic energies and the single adiabatic ground-state energy do not provide sufficient information
to exactly infer the $O(n^2)$ diabatic couplings. 
To address this problem, one can turn to developments in multistate EVB approaches~\cite{Vuilleumier1998,Schmitt1999,Day2002}.
For example, one crude solution would be to only consider the two lowest diabatic states at a time,
over the course of a molecular dynamics trajectory. Another would be to calculate all pairwise couplings in isolation. 
Lastly, one could use simple functional forms for the diabatic couplings, 
as commonly done in EVB models, or train a model that is more expressive and/or environmentally dependent.
Work along all of these lines is currently in progress.

Finally, we speculate that the CQEq framework could be used to address the
unphysical long-range CT problem of conventional, ground-state QEq. For example,
one could rescale the calculated diabatic couplings so that they vanish at large 
separations, guaranteeing that molecular complexes dissociate correctly into 
diabatic states without modifying their equilibrium description. Such an approach
would provide a smooth connection to QEq modifications that prohibit CT between predefined
fragments or molecules~\cite{Rick1994,Shimizu2004}.

\label{sec:conc}

\section*{Acknowledgements}
We thank Garvit Agarwal (Schr\"odinger Inc.)~for providing the nudged elastic
band geometries of EC on lithium.  This work was supported by the Columbia
Center for Computational Electrochemistry.  We acknowledge computing resources
from Columbia University’s Shared Research Computing Facility project, which is
supported by NIH Research Facility Improvement Grant 1G20RR030893-01, and
associated funds from the New York State Empire State Development, Division of
Science Technology and Innovation (NYSTAR) Contract C090171, both awarded April
15, 2010.  The Flatiron Institute is a division of the Simons Foundation.

\vspace{4em}

\section*{Data Availability Statement}
The data that support the findings of this study are available from the corresponding author upon reasonable request.

\newpage
\widetext
\begin{center}
    \textbf{Supporting Information for: Diabatic states of charge transfer with constrained charge equilibration}
\end{center}
\setcounter{section}{0}
\setcounter{equation}{0}
\setcounter{figure}{0}
\setcounter{table}{0}
\setcounter{page}{1}
\makeatletter
\renewcommand{\theequation}{S\arabic{equation}}
\renewcommand{\thesection}{S\arabic{section}}
\renewcommand{\thefigure}{S\arabic{figure}}

\section{Diatomic Molecule}
\label{app:diatomic}
Consider a diatomic molecule AB with bond length $R$. 
Enforcing overall charge-neutrality for the molecule (i.e., $q=q_\mathrm{A}=-q_\mathrm{B}$)
Eq.~(\ref{eq:qeq}) simplifies as 
\begin{equation}
\label{eq:qeqdiatomic}
\begin{split}
U_\mathrm{elec}(R,q) 
    &= q \left[\chi_\mathrm{A}-\chi_\mathrm{B}\right] + \frac{1}{2} q^2 \left[J_\mathrm{AA} + J_\mathrm{BB} 
     - 2J_\mathrm{AB}(R)\right]
\end{split}
\end{equation}
Defining $q_\mathrm{min}$ as the charge that minimizes $U_\mathrm{elec}$, we find
\begin{equation}
\label{eq:qmin}
    q_\mathrm{min}(R) = \left[\chi_\mathrm{B}-\chi_\mathrm{A}\right]/
    \left[J_\mathrm{AA}+J_\mathrm{BB}-2J_\mathrm{AB}(R)\right],
\end{equation}
and the corresponding electrostatic energy
\begin{equation}
\label{eq:Umin}
    U_\mathrm{elec,min} (R) =  -\frac{1}{2}\frac{\left[\chi_\mathrm{B}-\chi_\mathrm{A}\right]^2}
    {\left[J_\mathrm{AA}+J_\mathrm{BB}-2J_\mathrm{AB}(R)\right]}.
\end{equation}
Note, in Eqs.~(\ref{eq:qmin}) and (\ref{eq:Umin}), the $R$-dependence is entirely in the Coulomb term. 
In the limit of $R\to\infty$, the Coloumb integral vanishes such that 
\begin{equation}
    q_\mathrm{min}(R \to \infty) = \frac{\chi_\mathrm{B}-\chi_\mathrm{A}}{J_\mathrm{AA}+J_\mathrm{BB}}.
\end{equation}
Since $\chi_B\neq\chi_A$, this scheme predicts non-integer charge transfer at 
the dissociation limit. This error is well-known in the QEq literature
and can be alternatively seen from Eq.~(\ref{eq:Umin}) where $U_\mathrm{elec, min} (R \to \infty)$ is always less 
than zero (and hence less than the energy associated with neutral fragments).

\begin{figure}
    \includegraphics[scale=0.8]{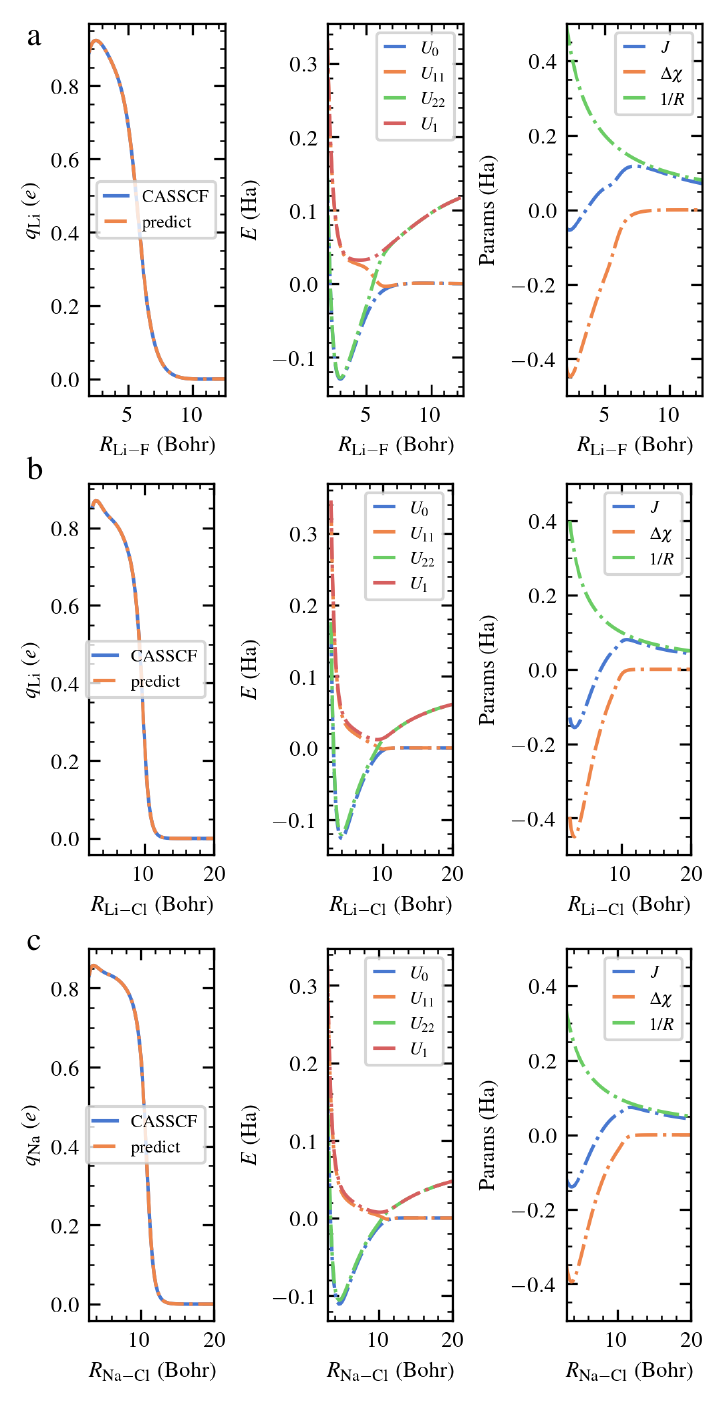}
    \caption[\centering]{Constrained QEq within the exact model applied to (a) LiF, (b) LiCl and (c) NaCl.} 
    \label{fig:diatom}
 \end{figure}

From Eq.~(\ref{eq:qmin}), it is clear that to achieve the correct dissociation limit, 
$\chi_\mathrm{A}$ and $\chi_\mathrm{B}$ need to be $R$-dependent such that 
their difference vanishes at infinite separation, such as might be achieved with
environment-aware electronegativities. 
For a diatomic problem, we can construct an exact model for illustration.
Noting that we only need the difference in electronegativities, we parameterize
\begin{equation}
\label{eq:Kparam}
    \Delta\chi(R) = \chi_\mathrm{A}(R)-\chi_\mathrm{B}(R) = \tilde{q}(R)\left[2J_\mathrm{AB}(R)-J_\mathrm{AA}-J_\mathrm{BB}\right],
\end{equation}
where $\tilde{q}(R)$ is the charge obtained from a reference quantum chemistry calculation. 
Equation~(\ref{eq:Kparam}) ensures that the reference charge $\tilde{q}$ minimizes
the electrostatic energy.
Next, one can calculate energies of the diabatic states 
\begin{subequations}
\begin{align}
    U_{11}(R) &= U_{0}(R) - \Delta\chi(R)\tilde{q}(R) 
        - \left[\frac{1}{2}J_\mathrm{AA}+\frac{1}{2}J_\mathrm{BB}-J_\mathrm{AB}(R)\right]\tilde{q}^2(R) \\
    U_{22}(R) &= U_{11}(R) + \Delta\chi(R) 
        + \frac{1}{2}J_\mathrm{AA}+\frac{1}{2}J_\mathrm{BB}-J_\mathrm{AB}(R).
\end{align}
\end{subequations}
The energy of the adiabatic excited state is given by
\begin{equation}
\label{eq:U1QEq}
    U_{1}(R) = U_{0}(R) + \left[\frac{1}{2}J_\mathrm{AA} 
        + \frac{1}{2}J_\mathrm{BB}-J_\mathrm{AB}(R)\right]\left[2\tilde{q}^2(R)-2\tilde{q}(R)+1\right].
\end{equation}
For both $\tilde{q}=0$ and $\tilde{q}=1$, this expression simplifies to
\begin{equation}
    U_{1}(R) = U_{0}(R) + \left[\frac{1}{2}J_\mathrm{AA} + \frac{1}{2}J_\mathrm{BB}-J_\mathrm{AB}(R)\right],
\end{equation}
such that the excitation energy at $R\rightarrow\infty$ fixes the sum of the
hardness parameters in terms of the atomic ionization potentials and electron
affinities 
\begin{equation}
   \frac{1}{2}J_\mathrm{AA}+\frac{1}{2}J_\mathrm{BB} = \mathrm{IP_A}+\mathrm{EA_B}.
\end{equation}
In principle, the typical QEq Coulomb energy $J_{AB}(R)$ can be used in Eq.~(\ref{eq:U1QEq}),
but we find that it gives a poor prediction of the adiabatic excitation energy.
For insight, we can determine the exact $J_\mathrm{AB}(R)$ needed to produce the exact
adiabatic excitation energy via
\begin{equation}
\begin{split}
    J_\mathrm{AB}(R) = \frac{1}{2}J_\mathrm{AA} + \frac{1}{2}J_\mathrm{BB}
                       - \frac{U_1(R)-U_0(R)}{2\tilde{q}^2(R)-2\tilde{q}(R)+1}
\end{split}
\end{equation}
where $U_1(R)-U_0(R)$ is obtained from a reference quantum chemistry calculation.

Because dissociation of diatomic molecules into open-shell species requires a multiconfigurational
treatment, we use CASSCF as our reference method. In Fig.~\ref{fig:diatom}, we show results
for LiF, LiCl, and NaCl, using a minimal (2,2) state-averaged CASSCF.
We emphasize that the purpose of this diatomic study is illustration only: the charges
and adiabatic energies are exactly predicted by CQEq by construction. The main results are 
in the rightmost panels, where the behavior of the exact $\Delta \chi(R)$ and $J_{AB}(R)$
are shown. As expected, the exact $\Delta \chi(R)$ goes to zero are large $R$ to enforce
the correct dissociation limit of neutral, open-shell atoms. The exact $J_{AB}(R)$ shows
the expected $1/R$ decay at large $R$, but it becomes negative at small $R$, i.e., the
interaction between the opposite charges ($q_\mathrm{A} = -q_\mathrm{B}$) 
is repulsive rather than attractive.
Perhaps it is unsurprising that at small bond lengths, a purely electrostatic treatment of the
electronic energy is not sufficient.

\section{Charge-Equilibration Parameters} 
The parameters that we use for ground state QEq, i.e., hardness $J_{ii}$ and exponents
$\zeta_{i}$, are the ones reported in Refs.~\onlinecite{Rappe1991} and
\onlinecite{Caldeweyher2019}, respectively. We have reproduced these values in Table
\ref{tab:params}. We find the best performance using this mixed set of parameters, although
the results obtained using any combination of parameters from these two works are qualitatively
similar.

\begin{table}[t]
    \centering
    \begin{tabular}{lcc}
    \hline\hline
    Element & $J_{ii}$ (a.u.) & $\zeta$ (a.u.)\\
    \hline
    H  & 1.0964 & 1.0698\\
    Li & 0.9742 & 0.4174\\
    C  & 0.6165 & 0.8563\\
    N  & 0.6565 & 0.9089\\
    O  & 0.6793 & 0.9745\\
    \hline\hline
    \end{tabular}
    \caption{Atomic hardness parameters (from reference \onlinecite{Rappe1991}) and Slater coefficients for the highest occupied ns orbital (from reference \onlinecite{Caldeweyher2019}) that are used to parametrize the Coulomb matrix.
    }
    \label{tab:params}
    \end{table}

 \section{Electronegativity for anthracene-TCNE}

\begin{figure}[t]
    \includegraphics[scale=0.8]{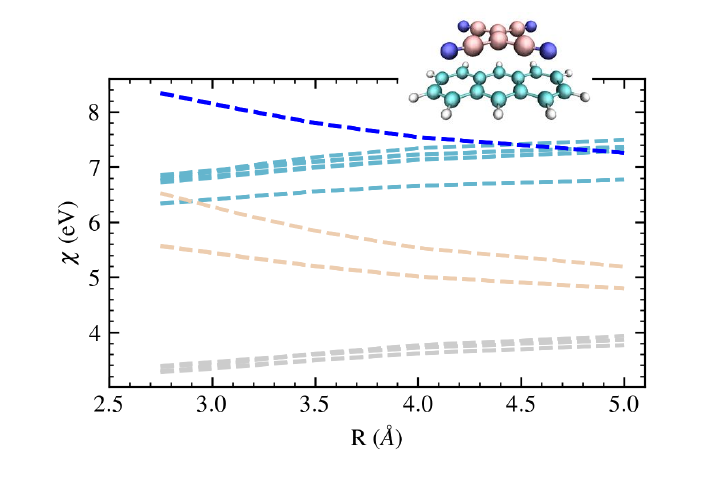}
    \caption[\centering]{(a) Atomic electronegativities calculated using
Eq.~(\ref{eq:chiR}) for anthracene-tetracyanoethylene complex 
} 
 \label{fig:chiqCTSI}
\end{figure}

In Figure \ref{fig:chiqCTSI}, we show atomic electronegativity values for the anthracene-TCNE complex. The color coding of curves follows those in the molecular graphic: nitrogen (blue), hydrogen (grey), anthracene carbon (cyan), 
and TCNE carbon (light brown). 

\vspace{4em}


\section*{References}
\begin{thebibliography}{57}%
\makeatletter
\providecommand \@ifxundefined [1]{%
 \@ifx{#1\undefined}
}%
\providecommand \@ifnum [1]{%
 \ifnum #1\expandafter \@firstoftwo
 \else \expandafter \@secondoftwo
 \fi
}%
\providecommand \@ifx [1]{%
 \ifx #1\expandafter \@firstoftwo
 \else \expandafter \@secondoftwo
 \fi
}%
\providecommand \natexlab [1]{#1}%
\providecommand \enquote  [1]{``#1''}%
\providecommand \bibnamefont  [1]{#1}%
\providecommand \bibfnamefont [1]{#1}%
\providecommand \citenamefont [1]{#1}%
\providecommand \href@noop [0]{\@secondoftwo}%
\providecommand \href [0]{\begingroup \@sanitize@url \@href}%
\providecommand \@href[1]{\@@startlink{#1}\@@href}%
\providecommand \@@href[1]{\endgroup#1\@@endlink}%
\providecommand \@sanitize@url [0]{\catcode `\\12\catcode `\$12\catcode `\&12\catcode `\#12\catcode `\^12\catcode `\_12\catcode `\%12\relax}%
\providecommand \@@startlink[1]{}%
\providecommand \@@endlink[0]{}%
\providecommand \url  [0]{\begingroup\@sanitize@url \@url }%
\providecommand \@url [1]{\endgroup\@href {#1}{\urlprefix }}%
\providecommand \urlprefix  [0]{URL }%
\providecommand \Eprint [0]{\href }%
\providecommand \doibase [0]{https://doi.org/}%
\providecommand \selectlanguage [0]{\@gobble}%
\providecommand \bibinfo  [0]{\@secondoftwo}%
\providecommand \bibfield  [0]{\@secondoftwo}%
\providecommand \translation [1]{[#1]}%
\providecommand \BibitemOpen [0]{}%
\providecommand \bibitemStop [0]{}%
\providecommand \bibitemNoStop [0]{.\EOS\space}%
\providecommand \EOS [0]{\spacefactor3000\relax}%
\providecommand \BibitemShut  [1]{\csname bibitem#1\endcsname}%
\let\auto@bib@innerbib\@empty
\bibitem [{\citenamefont {Rappe}\ and\ \citenamefont {Goddard}(1991)}]{Rappe1991}%
  \BibitemOpen
  \bibfield  {author} {\bibinfo {author} {\bibfnamefont {A.~K.}\ \bibnamefont {Rappe}}\ and\ \bibinfo {author} {\bibfnamefont {W.~A.}\ \bibnamefont {Goddard}},\ }\bibfield  {title} {\enquote {\bibinfo {title} {Charge equilibration for molecular dynamics simulations},}\ }\href {https://doi.org/10.1021/j100161a070} {\bibfield  {journal} {\bibinfo  {journal} {J. Phys. Chem.}\ }\textbf {\bibinfo {volume} {95}},\ \bibinfo {pages} {3358--3363} (\bibinfo {year} {1991})}\BibitemShut {NoStop}%
\bibitem [{\citenamefont {Sanderson}(1976)}]{Sanderson1976}%
  \BibitemOpen
  \bibfield  {author} {\bibinfo {author} {\bibfnamefont {R.~T.}\ \bibnamefont {Sanderson}},\ }\href@noop {} {\emph {\bibinfo {title} {Chemical bonds and bond energy}}},\ \bibinfo {edition} {2nd}\ ed.,\ \bibinfo {series} {Physical chemistry, a series of monographs}\ No.\ \bibinfo {number} {v. 21}\ (\bibinfo  {publisher} {Academic Press},\ \bibinfo {address} {New York},\ \bibinfo {year} {1976})\ \bibinfo {note} {includes bibliographical references and index}\BibitemShut {NoStop}%
\bibitem [{\citenamefont {Mortier}, \citenamefont {Ghosh},\ and\ \citenamefont {Shankar}(1986)}]{Mortier1986}%
  \BibitemOpen
  \bibfield  {author} {\bibinfo {author} {\bibfnamefont {W.~J.}\ \bibnamefont {Mortier}}, \bibinfo {author} {\bibfnamefont {S.~K.}\ \bibnamefont {Ghosh}},\ and\ \bibinfo {author} {\bibfnamefont {S.}~\bibnamefont {Shankar}},\ }\bibfield  {title} {\enquote {\bibinfo {title} {Electronegativity-equalization method for the calculation of atomic charges in molecules},}\ }\href {https://doi.org/10.1021/ja00275a013} {\bibfield  {journal} {\bibinfo  {journal} {J. Am. Chem. Soc.}\ }\textbf {\bibinfo {volume} {108}},\ \bibinfo {pages} {4315--4320} (\bibinfo {year} {1986})}\BibitemShut {NoStop}%
\bibitem [{\citenamefont {Rick}\ and\ \citenamefont {Berne}(1996)}]{Rick1996}%
  \BibitemOpen
  \bibfield  {author} {\bibinfo {author} {\bibfnamefont {S.~W.}\ \bibnamefont {Rick}}\ and\ \bibinfo {author} {\bibfnamefont {B.~J.}\ \bibnamefont {Berne}},\ }\bibfield  {title} {\enquote {\bibinfo {title} {Dynamical fluctuating charge force fields: The aqueous solvation of amides},}\ }\href {https://doi.org/10.1021/ja952535b} {\bibfield  {journal} {\bibinfo  {journal} {J. Am. Chem. Soc.}\ }\textbf {\bibinfo {volume} {118}},\ \bibinfo {pages} {672--679} (\bibinfo {year} {1996})}\BibitemShut {NoStop}%
\bibitem [{\citenamefont {Patel}, \citenamefont {Mackerell},\ and\ \citenamefont {Brooks}(2004)}]{Patel2004}%
  \BibitemOpen
  \bibfield  {author} {\bibinfo {author} {\bibfnamefont {S.}~\bibnamefont {Patel}}, \bibinfo {author} {\bibfnamefont {A.~D.}\ \bibnamefont {Mackerell}},\ and\ \bibinfo {author} {\bibfnamefont {C.~L.}\ \bibnamefont {Brooks}},\ }\bibfield  {title} {\enquote {\bibinfo {title} {{CHARMM} fluctuating charge force field for proteins: Ii protein/solvent properties from molecular dynamics simulations using a nonadditive electrostatic model},}\ }\href {https://doi.org/10.1002/jcc.20077} {\bibfield  {journal} {\bibinfo  {journal} {J. Comput. Chem.}\ }\textbf {\bibinfo {volume} {25}},\ \bibinfo {pages} {1504--1514} (\bibinfo {year} {2004})}\BibitemShut {NoStop}%
\bibitem [{\citenamefont {Patel}\ and\ \citenamefont {Brooks}(2006)}]{Patel2006}%
  \BibitemOpen
  \bibfield  {author} {\bibinfo {author} {\bibfnamefont {S.}~\bibnamefont {Patel}}\ and\ \bibinfo {author} {\bibfnamefont {C.~L.}\ \bibnamefont {Brooks}},\ }\bibfield  {title} {\enquote {\bibinfo {title} {Fluctuating charge force fields: recent developments and applications from small molecules to macromolecular biological systems},}\ }\href {https://doi.org/10.1080/08927020600726708} {\bibfield  {journal} {\bibinfo  {journal} {Mol. Simulat.}\ }\textbf {\bibinfo {volume} {32}},\ \bibinfo {pages} {231--249} (\bibinfo {year} {2006})}\BibitemShut {NoStop}%
\bibitem [{\citenamefont {Jing}\ \emph {et~al.}(2019)\citenamefont {Jing}, \citenamefont {Liu}, \citenamefont {Cheng}, \citenamefont {Qi}, \citenamefont {Walker}, \citenamefont {Piquemal},\ and\ \citenamefont {Ren}}]{Jing2019}%
  \BibitemOpen
  \bibfield  {author} {\bibinfo {author} {\bibfnamefont {Z.}~\bibnamefont {Jing}}, \bibinfo {author} {\bibfnamefont {C.}~\bibnamefont {Liu}}, \bibinfo {author} {\bibfnamefont {S.~Y.}\ \bibnamefont {Cheng}}, \bibinfo {author} {\bibfnamefont {R.}~\bibnamefont {Qi}}, \bibinfo {author} {\bibfnamefont {B.~D.}\ \bibnamefont {Walker}}, \bibinfo {author} {\bibfnamefont {J.-P.}\ \bibnamefont {Piquemal}},\ and\ \bibinfo {author} {\bibfnamefont {P.}~\bibnamefont {Ren}},\ }\bibfield  {title} {\enquote {\bibinfo {title} {Polarizable force fields for biomolecular simulations: Recent advances and applications},}\ }\href {https://doi.org/10.1146/annurev-biophys-070317-033349} {\bibfield  {journal} {\bibinfo  {journal} {Annu. Rev. Biophys.}\ }\textbf {\bibinfo {volume} {48}},\ \bibinfo {pages} {371--394} (\bibinfo {year} {2019})}\BibitemShut {NoStop}%
\bibitem [{\citenamefont {Caldeweyher}\ \emph {et~al.}(2019)\citenamefont {Caldeweyher}, \citenamefont {Ehlert}, \citenamefont {Hansen}, \citenamefont {Neugebauer}, \citenamefont {Spicher}, \citenamefont {Bannwarth},\ and\ \citenamefont {Grimme}}]{Caldeweyher2019}%
  \BibitemOpen
  \bibfield  {author} {\bibinfo {author} {\bibfnamefont {E.}~\bibnamefont {Caldeweyher}}, \bibinfo {author} {\bibfnamefont {S.}~\bibnamefont {Ehlert}}, \bibinfo {author} {\bibfnamefont {A.}~\bibnamefont {Hansen}}, \bibinfo {author} {\bibfnamefont {H.}~\bibnamefont {Neugebauer}}, \bibinfo {author} {\bibfnamefont {S.}~\bibnamefont {Spicher}}, \bibinfo {author} {\bibfnamefont {C.}~\bibnamefont {Bannwarth}},\ and\ \bibinfo {author} {\bibfnamefont {S.}~\bibnamefont {Grimme}},\ }\bibfield  {title} {\enquote {\bibinfo {title} {A generally applicable atomic-charge dependent {London} dispersion correction},}\ }\href {https://doi.org/10.1063/1.5090222} {\bibfield  {journal} {\bibinfo  {journal} {J. Chem. Phys.}\ }\textbf {\bibinfo {volume} {150}},\ \bibinfo {pages} {154122} (\bibinfo {year} {2019})}\BibitemShut {NoStop}%
\bibitem [{\citenamefont {van Duin}\ \emph {et~al.}(2001)\citenamefont {van Duin}, \citenamefont {Dasgupta}, \citenamefont {Lorant},\ and\ \citenamefont {Goddard}}]{Duin2001}%
  \BibitemOpen
  \bibfield  {author} {\bibinfo {author} {\bibfnamefont {A.~C.~T.}\ \bibnamefont {van Duin}}, \bibinfo {author} {\bibfnamefont {S.}~\bibnamefont {Dasgupta}}, \bibinfo {author} {\bibfnamefont {F.}~\bibnamefont {Lorant}},\ and\ \bibinfo {author} {\bibfnamefont {W.~A.}\ \bibnamefont {Goddard}},\ }\bibfield  {title} {\enquote {\bibinfo {title} {{ReaxFF}: A reactive force field for hydrocarbons},}\ }\href {https://doi.org/10.1021/jp004368u} {\bibfield  {journal} {\bibinfo  {journal} {J. Phys. Chem. A}\ }\textbf {\bibinfo {volume} {105}},\ \bibinfo {pages} {9396--9409} (\bibinfo {year} {2001})}\BibitemShut {NoStop}%
\bibitem [{\citenamefont {Senftle}\ \emph {et~al.}(2016)\citenamefont {Senftle}, \citenamefont {Hong}, \citenamefont {Islam}, \citenamefont {Kylasa}, \citenamefont {Zheng}, \citenamefont {Shin}, \citenamefont {Junkermeier}, \citenamefont {Engel-Herbert}, \citenamefont {Janik}, \citenamefont {Aktulga}, \citenamefont {Verstraelen}, \citenamefont {Grama},\ and\ \citenamefont {van Duin}}]{Senftle2016}%
  \BibitemOpen
  \bibfield  {author} {\bibinfo {author} {\bibfnamefont {T.~P.}\ \bibnamefont {Senftle}}, \bibinfo {author} {\bibfnamefont {S.}~\bibnamefont {Hong}}, \bibinfo {author} {\bibfnamefont {M.~M.}\ \bibnamefont {Islam}}, \bibinfo {author} {\bibfnamefont {S.~B.}\ \bibnamefont {Kylasa}}, \bibinfo {author} {\bibfnamefont {Y.}~\bibnamefont {Zheng}}, \bibinfo {author} {\bibfnamefont {Y.~K.}\ \bibnamefont {Shin}}, \bibinfo {author} {\bibfnamefont {C.}~\bibnamefont {Junkermeier}}, \bibinfo {author} {\bibfnamefont {R.}~\bibnamefont {Engel-Herbert}}, \bibinfo {author} {\bibfnamefont {M.~J.}\ \bibnamefont {Janik}}, \bibinfo {author} {\bibfnamefont {H.~M.}\ \bibnamefont {Aktulga}}, \bibinfo {author} {\bibfnamefont {T.}~\bibnamefont {Verstraelen}}, \bibinfo {author} {\bibfnamefont {A.}~\bibnamefont {Grama}},\ and\ \bibinfo {author} {\bibfnamefont {A.~C.~T.}\ \bibnamefont {van Duin}},\ }\bibfield  {title} {\enquote {\bibinfo {title} {The {ReaxFF} reactive force-field: development, applications and future directions},}\ }\href
  {https://doi.org/10.1038/npjcompumats.2015.11} {\bibfield  {journal} {\bibinfo  {journal} {npj Comput. Mater.}\ }\textbf {\bibinfo {volume} {2}},\ \bibinfo {pages} {1--14} (\bibinfo {year} {2016})}\BibitemShut {NoStop}%
\bibitem [{\citenamefont {Li}\ and\ \citenamefont {Voth}(2021)}]{Li2021}%
  \BibitemOpen
  \bibfield  {author} {\bibinfo {author} {\bibfnamefont {C.}~\bibnamefont {Li}}\ and\ \bibinfo {author} {\bibfnamefont {G.~A.}\ \bibnamefont {Voth}},\ }\bibfield  {title} {\enquote {\bibinfo {title} {Accurate and transferable reactive molecular dynamics models from constrained density functional theory},}\ }\href {https://doi.org/10.1021/acs.jpcb.1c05992} {\bibfield  {journal} {\bibinfo  {journal} {J. Phys. Chem. B}\ }\textbf {\bibinfo {volume} {125}},\ \bibinfo {pages} {10471--10480} (\bibinfo {year} {2021})}\BibitemShut {NoStop}%
\bibitem [{\citenamefont {Behler}(2021)}]{Behler2021}%
  \BibitemOpen
  \bibfield  {author} {\bibinfo {author} {\bibfnamefont {J.}~\bibnamefont {Behler}},\ }\bibfield  {title} {\enquote {\bibinfo {title} {Four generations of high-dimensional neural network potentials},}\ }\href {https://doi.org/10.1021/acs.chemrev.0c00868} {\bibfield  {journal} {\bibinfo  {journal} {Chem. Rev.}\ }\textbf {\bibinfo {volume} {121}},\ \bibinfo {pages} {10037--10072} (\bibinfo {year} {2021})}\BibitemShut {NoStop}%
\bibitem [{\citenamefont {Unke}\ \emph {et~al.}(2021)\citenamefont {Unke}, \citenamefont {Chmiela}, \citenamefont {Sauceda}, \citenamefont {Gastegger}, \citenamefont {Poltavsky}, \citenamefont {Schütt}, \citenamefont {Tkatchenko},\ and\ \citenamefont {Müller}}]{Unke2021}%
  \BibitemOpen
  \bibfield  {author} {\bibinfo {author} {\bibfnamefont {O.~T.}\ \bibnamefont {Unke}}, \bibinfo {author} {\bibfnamefont {S.}~\bibnamefont {Chmiela}}, \bibinfo {author} {\bibfnamefont {H.~E.}\ \bibnamefont {Sauceda}}, \bibinfo {author} {\bibfnamefont {M.}~\bibnamefont {Gastegger}}, \bibinfo {author} {\bibfnamefont {I.}~\bibnamefont {Poltavsky}}, \bibinfo {author} {\bibfnamefont {K.~T.}\ \bibnamefont {Schütt}}, \bibinfo {author} {\bibfnamefont {A.}~\bibnamefont {Tkatchenko}},\ and\ \bibinfo {author} {\bibfnamefont {K.-R.}\ \bibnamefont {Müller}},\ }\bibfield  {title} {\enquote {\bibinfo {title} {Machine learning force fields},}\ }\href {https://doi.org/10.1021/acs.chemrev.0c01111} {\bibfield  {journal} {\bibinfo  {journal} {Chem. Rev.}\ }\textbf {\bibinfo {volume} {121}},\ \bibinfo {pages} {10142--10186} (\bibinfo {year} {2021})}\BibitemShut {NoStop}%
\bibitem [{\citenamefont {Ko}\ \emph {et~al.}(2021)\citenamefont {Ko}, \citenamefont {Finkler}, \citenamefont {Goedecker},\ and\ \citenamefont {Behler}}]{Ko2021}%
  \BibitemOpen
  \bibfield  {author} {\bibinfo {author} {\bibfnamefont {T.~W.}\ \bibnamefont {Ko}}, \bibinfo {author} {\bibfnamefont {J.~A.}\ \bibnamefont {Finkler}}, \bibinfo {author} {\bibfnamefont {S.}~\bibnamefont {Goedecker}},\ and\ \bibinfo {author} {\bibfnamefont {J.}~\bibnamefont {Behler}},\ }\bibfield  {title} {\enquote {\bibinfo {title} {A fourth-generation high-dimensional neural network potential with accurate electrostatics including non-local charge transfer},}\ }\href {https://doi.org/10.1038/s41467-020-20427-2} {\bibfield  {journal} {\bibinfo  {journal} {Nat. Commun.}\ }\textbf {\bibinfo {volume} {12}},\ \bibinfo {pages} {398} (\bibinfo {year} {2021})}\BibitemShut {NoStop}%
\bibitem [{\citenamefont {Deng}\ \emph {et~al.}(2023)\citenamefont {Deng}, \citenamefont {Zhong}, \citenamefont {Jun}, \citenamefont {Riebesell}, \citenamefont {Han}, \citenamefont {Bartel},\ and\ \citenamefont {Ceder}}]{Deng2023}%
  \BibitemOpen
  \bibfield  {author} {\bibinfo {author} {\bibfnamefont {B.}~\bibnamefont {Deng}}, \bibinfo {author} {\bibfnamefont {P.}~\bibnamefont {Zhong}}, \bibinfo {author} {\bibfnamefont {K.}~\bibnamefont {Jun}}, \bibinfo {author} {\bibfnamefont {J.}~\bibnamefont {Riebesell}}, \bibinfo {author} {\bibfnamefont {K.}~\bibnamefont {Han}}, \bibinfo {author} {\bibfnamefont {C.~J.}\ \bibnamefont {Bartel}},\ and\ \bibinfo {author} {\bibfnamefont {G.}~\bibnamefont {Ceder}},\ }\bibfield  {title} {\enquote {\bibinfo {title} {{CHGNet} as a pretrained universal neural network potential for charge-informed atomistic modelling},}\ }\href {https://doi.org/10.1038/s42256-023-00716-3} {\bibfield  {journal} {\bibinfo  {journal} {Nat. Mach. Intell.}\ }\textbf {\bibinfo {volume} {5}},\ \bibinfo {pages} {1031--1041} (\bibinfo {year} {2023})}\BibitemShut {NoStop}%
\bibitem [{\citenamefont {Kovács}\ \emph {et~al.}(2023)\citenamefont {Kovács}, \citenamefont {Batatia}, \citenamefont {Arany},\ and\ \citenamefont {Csányi}}]{Kovacs2023}%
  \BibitemOpen
  \bibfield  {author} {\bibinfo {author} {\bibfnamefont {D.~P.}\ \bibnamefont {Kovács}}, \bibinfo {author} {\bibfnamefont {I.}~\bibnamefont {Batatia}}, \bibinfo {author} {\bibfnamefont {E.~S.}\ \bibnamefont {Arany}},\ and\ \bibinfo {author} {\bibfnamefont {G.}~\bibnamefont {Csányi}},\ }\bibfield  {title} {\enquote {\bibinfo {title} {Evaluation of the {MACE} force field architecture: From medicinal chemistry to materials science},}\ }\href {https://doi.org/10.1063/5.0155322} {\bibfield  {journal} {\bibinfo  {journal} {J Chem Phys}\ }\textbf {\bibinfo {volume} {159}},\ \bibinfo {pages} {044118} (\bibinfo {year} {2023})}\BibitemShut {NoStop}%
\bibitem [{\citenamefont {Chen}\ \emph {et~al.}(2018)\citenamefont {Chen}, \citenamefont {Liu}, \citenamefont {Fang}, \citenamefont {Dral},\ and\ \citenamefont {Cui}}]{Chen2018}%
  \BibitemOpen
  \bibfield  {author} {\bibinfo {author} {\bibfnamefont {W.-K.}\ \bibnamefont {Chen}}, \bibinfo {author} {\bibfnamefont {X.-Y.}\ \bibnamefont {Liu}}, \bibinfo {author} {\bibfnamefont {W.-H.}\ \bibnamefont {Fang}}, \bibinfo {author} {\bibfnamefont {P.~O.}\ \bibnamefont {Dral}},\ and\ \bibinfo {author} {\bibfnamefont {G.}~\bibnamefont {Cui}},\ }\bibfield  {title} {\enquote {\bibinfo {title} {Deep learning for nonadiabatic excited-state dynamics},}\ }\href {https://doi.org/10.1021/acs.jpclett.8b03026} {\bibfield  {journal} {\bibinfo  {journal} {J. Phys. Chem. Lett.}\ }\textbf {\bibinfo {volume} {9}},\ \bibinfo {pages} {6702--6708} (\bibinfo {year} {2018})}\BibitemShut {NoStop}%
\bibitem [{\citenamefont {Dral}, \citenamefont {Barbatti},\ and\ \citenamefont {Thiel}(2018)}]{Dral2018}%
  \BibitemOpen
  \bibfield  {author} {\bibinfo {author} {\bibfnamefont {P.~O.}\ \bibnamefont {Dral}}, \bibinfo {author} {\bibfnamefont {M.}~\bibnamefont {Barbatti}},\ and\ \bibinfo {author} {\bibfnamefont {W.}~\bibnamefont {Thiel}},\ }\bibfield  {title} {\enquote {\bibinfo {title} {Nonadiabatic excited-state dynamics with machine learning},}\ }\href {https://doi.org/10.1021/acs.jpclett.8b02469} {\bibfield  {journal} {\bibinfo  {journal} {J. Phys. Chem. Lett.}\ }\textbf {\bibinfo {volume} {9}},\ \bibinfo {pages} {5660--5663} (\bibinfo {year} {2018})}\BibitemShut {NoStop}%
\bibitem [{\citenamefont {Westermayr}, \citenamefont {Gastegger},\ and\ \citenamefont {Marquetand}(2020)}]{Westermayr2020}%
  \BibitemOpen
  \bibfield  {author} {\bibinfo {author} {\bibfnamefont {J.}~\bibnamefont {Westermayr}}, \bibinfo {author} {\bibfnamefont {M.}~\bibnamefont {Gastegger}},\ and\ \bibinfo {author} {\bibfnamefont {P.}~\bibnamefont {Marquetand}},\ }\bibfield  {title} {\enquote {\bibinfo {title} {Combining {SchNet} and {SHARC}: The {SchNarc} machine learning approach for excited-state dynamics},}\ }\href {https://doi.org/10.1021/acs.jpclett.0c00527} {\bibfield  {journal} {\bibinfo  {journal} {J. Phys. Chem. Lett.}\ }\textbf {\bibinfo {volume} {11}},\ \bibinfo {pages} {3828--3834} (\bibinfo {year} {2020})}\BibitemShut {NoStop}%
\bibitem [{\citenamefont {Dupuy}\ and\ \citenamefont {Maitra}(2024)}]{Dupuy2024}%
  \BibitemOpen
  \bibfield  {author} {\bibinfo {author} {\bibfnamefont {L.}~\bibnamefont {Dupuy}}\ and\ \bibinfo {author} {\bibfnamefont {N.~T.}\ \bibnamefont {Maitra}},\ }\href {https://doi.org/10.48550/ARXIV.2407.10881} {\enquote {\bibinfo {title} {Exciting {DeePMD}: Learning excited state energies, forces, and non-adiabatic couplings},}\ } (\bibinfo {year} {2024}),\ \Eprint {https://arxiv.org/abs/2407.10881} {arXiv:2407.10881} \BibitemShut {NoStop}%
\bibitem [{\citenamefont {Richardson}(2023)}]{Richardson2023}%
  \BibitemOpen
  \bibfield  {author} {\bibinfo {author} {\bibfnamefont {J.~O.}\ \bibnamefont {Richardson}},\ }\bibfield  {title} {\enquote {\bibinfo {title} {Machine learning of double-valued nonadiabatic coupling vectors around conical intersections},}\ }\href {https://doi.org/10.1063/5.0133191} {\bibfield  {journal} {\bibinfo  {journal} {J. Chem. Phys.}\ }\textbf {\bibinfo {volume} {158}},\ \bibinfo {pages} {011102} (\bibinfo {year} {2023})}\BibitemShut {NoStop}%
\bibitem [{\citenamefont {Kaduk}, \citenamefont {Kowalczyk},\ and\ \citenamefont {Van~Voorhis}(2011)}]{Kaduk2011}%
  \BibitemOpen
  \bibfield  {author} {\bibinfo {author} {\bibfnamefont {B.}~\bibnamefont {Kaduk}}, \bibinfo {author} {\bibfnamefont {T.}~\bibnamefont {Kowalczyk}},\ and\ \bibinfo {author} {\bibfnamefont {T.}~\bibnamefont {Van~Voorhis}},\ }\bibfield  {title} {\enquote {\bibinfo {title} {Constrained density functional theory},}\ }\href {https://doi.org/10.1021/cr200148b} {\bibfield  {journal} {\bibinfo  {journal} {Chem. Rev.}\ }\textbf {\bibinfo {volume} {112}},\ \bibinfo {pages} {321--370} (\bibinfo {year} {2011})}\BibitemShut {NoStop}%
\bibitem [{\citenamefont {Kuharski}\ \emph {et~al.}(1988)\citenamefont {Kuharski}, \citenamefont {Bader}, \citenamefont {Chandler}, \citenamefont {Sprik}, \citenamefont {Klein},\ and\ \citenamefont {Impey}}]{Kuharski1988}%
  \BibitemOpen
  \bibfield  {author} {\bibinfo {author} {\bibfnamefont {R.~A.}\ \bibnamefont {Kuharski}}, \bibinfo {author} {\bibfnamefont {J.~S.}\ \bibnamefont {Bader}}, \bibinfo {author} {\bibfnamefont {D.}~\bibnamefont {Chandler}}, \bibinfo {author} {\bibfnamefont {M.}~\bibnamefont {Sprik}}, \bibinfo {author} {\bibfnamefont {M.~L.}\ \bibnamefont {Klein}},\ and\ \bibinfo {author} {\bibfnamefont {R.~W.}\ \bibnamefont {Impey}},\ }\bibfield  {title} {\enquote {\bibinfo {title} {Molecular model for aqueous ferrous–ferric electron transfer},}\ }\href {https://doi.org/10.1063/1.454929} {\bibfield  {journal} {\bibinfo  {journal} {J. Chem. Phys.}\ }\textbf {\bibinfo {volume} {89}},\ \bibinfo {pages} {3248--3257} (\bibinfo {year} {1988})}\BibitemShut {NoStop}%
\bibitem [{\citenamefont {Sit}, \citenamefont {Cococcioni},\ and\ \citenamefont {Marzari}(2006)}]{Sit2006}%
  \BibitemOpen
  \bibfield  {author} {\bibinfo {author} {\bibfnamefont {P.~H.-L.}\ \bibnamefont {Sit}}, \bibinfo {author} {\bibfnamefont {M.}~\bibnamefont {Cococcioni}},\ and\ \bibinfo {author} {\bibfnamefont {N.}~\bibnamefont {Marzari}},\ }\bibfield  {title} {\enquote {\bibinfo {title} {Realistic quantitative descriptions of electron transfer reactions: Diabatic free-energy surfaces from first-principles molecular dynamics},}\ }\href {https://doi.org/10.1103/physrevlett.97.028303} {\bibfield  {journal} {\bibinfo  {journal} {Phys. Rev. Lett.}\ }\textbf {\bibinfo {volume} {97}},\ \bibinfo {pages} {028303} (\bibinfo {year} {2006})}\BibitemShut {NoStop}%
\bibitem [{\citenamefont {Warshel}\ and\ \citenamefont {Weiss}(1980)}]{Warshel1980}%
  \BibitemOpen
  \bibfield  {author} {\bibinfo {author} {\bibfnamefont {A.}~\bibnamefont {Warshel}}\ and\ \bibinfo {author} {\bibfnamefont {R.~M.}\ \bibnamefont {Weiss}},\ }\bibfield  {title} {\enquote {\bibinfo {title} {An empirical valence bond approach for comparing reactions in solutions and in enzymes},}\ }\href {https://doi.org/10.1021/ja00540a008} {\bibfield  {journal} {\bibinfo  {journal} {J. Am. Chem. Soc.}\ }\textbf {\bibinfo {volume} {102}},\ \bibinfo {pages} {6218--6226} (\bibinfo {year} {1980})}\BibitemShut {NoStop}%
\bibitem [{\citenamefont {Kamerlin}\ and\ \citenamefont {Warshel}(2011)}]{Kamerlin2011}%
  \BibitemOpen
  \bibfield  {author} {\bibinfo {author} {\bibfnamefont {S.~C.~L.}\ \bibnamefont {Kamerlin}}\ and\ \bibinfo {author} {\bibfnamefont {A.}~\bibnamefont {Warshel}},\ }\bibfield  {title} {\enquote {\bibinfo {title} {The empirical valence bond model: theory and applications},}\ }\href {https://doi.org/10.1002/wcms.10} {\bibfield  {journal} {\bibinfo  {journal} {WIREs Comput. Mol. Sci.}\ }\textbf {\bibinfo {volume} {1}},\ \bibinfo {pages} {30--45} (\bibinfo {year} {2011})}\BibitemShut {NoStop}%
\bibitem [{\citenamefont {Kim}\ \emph {et~al.}(2000)\citenamefont {Kim}, \citenamefont {Corchado}, \citenamefont {Villà}, \citenamefont {Xing},\ and\ \citenamefont {Truhlar}}]{Kim2000}%
  \BibitemOpen
  \bibfield  {author} {\bibinfo {author} {\bibfnamefont {Y.}~\bibnamefont {Kim}}, \bibinfo {author} {\bibfnamefont {J.~C.}\ \bibnamefont {Corchado}}, \bibinfo {author} {\bibfnamefont {J.}~\bibnamefont {Villà}}, \bibinfo {author} {\bibfnamefont {J.}~\bibnamefont {Xing}},\ and\ \bibinfo {author} {\bibfnamefont {D.~G.}\ \bibnamefont {Truhlar}},\ }\bibfield  {title} {\enquote {\bibinfo {title} {Multiconfiguration molecular mechanics algorithm for potential energy surfaces of chemical reactions},}\ }\href {https://doi.org/10.1063/1.480846} {\bibfield  {journal} {\bibinfo  {journal} {J. Chem. Phys.}\ }\textbf {\bibinfo {volume} {112}},\ \bibinfo {pages} {2718--2735} (\bibinfo {year} {2000})}\BibitemShut {NoStop}%
\bibitem [{\citenamefont {Hong}, \citenamefont {Rosta},\ and\ \citenamefont {Warshel}(2006)}]{Hong2006}%
  \BibitemOpen
  \bibfield  {author} {\bibinfo {author} {\bibfnamefont {G.}~\bibnamefont {Hong}}, \bibinfo {author} {\bibfnamefont {E.}~\bibnamefont {Rosta}},\ and\ \bibinfo {author} {\bibfnamefont {A.}~\bibnamefont {Warshel}},\ }\bibfield  {title} {\enquote {\bibinfo {title} {Using the constrained {DFT} approach in generating diabatic surfaces and off diagonal empirical valence bond terms for modeling reactions in condensed phases},}\ }\href {https://doi.org/10.1021/jp0625199} {\bibfield  {journal} {\bibinfo  {journal} {J. Phys. Chem. B}\ }\textbf {\bibinfo {volume} {110}},\ \bibinfo {pages} {19570--19574} (\bibinfo {year} {2006})}\BibitemShut {NoStop}%
\bibitem [{\citenamefont {Ghasemi}\ \emph {et~al.}(2015)\citenamefont {Ghasemi}, \citenamefont {Hofstetter}, \citenamefont {Saha},\ and\ \citenamefont {Goedecker}}]{Ghasemi2015}%
  \BibitemOpen
  \bibfield  {author} {\bibinfo {author} {\bibfnamefont {S.~A.}\ \bibnamefont {Ghasemi}}, \bibinfo {author} {\bibfnamefont {A.}~\bibnamefont {Hofstetter}}, \bibinfo {author} {\bibfnamefont {S.}~\bibnamefont {Saha}},\ and\ \bibinfo {author} {\bibfnamefont {S.}~\bibnamefont {Goedecker}},\ }\bibfield  {title} {\enquote {\bibinfo {title} {Interatomic potentials for ionic systems with density functional accuracy based on charge densities obtained by a neural network},}\ }\href {https://doi.org/10.1103/physrevb.92.045131} {\bibfield  {journal} {\bibinfo  {journal} {Phys. Rev. B}\ }\textbf {\bibinfo {volume} {92}},\ \bibinfo {pages} {045131} (\bibinfo {year} {2015})}\BibitemShut {NoStop}%
\bibitem [{\citenamefont {Jacobson}\ \emph {et~al.}(2022)\citenamefont {Jacobson}, \citenamefont {Stevenson}, \citenamefont {Ramezanghorbani}, \citenamefont {Ghoreishi}, \citenamefont {Leswing}, \citenamefont {Harder},\ and\ \citenamefont {Abel}}]{Jacobson2022}%
  \BibitemOpen
  \bibfield  {author} {\bibinfo {author} {\bibfnamefont {L.~D.}\ \bibnamefont {Jacobson}}, \bibinfo {author} {\bibfnamefont {J.~M.}\ \bibnamefont {Stevenson}}, \bibinfo {author} {\bibfnamefont {F.}~\bibnamefont {Ramezanghorbani}}, \bibinfo {author} {\bibfnamefont {D.}~\bibnamefont {Ghoreishi}}, \bibinfo {author} {\bibfnamefont {K.}~\bibnamefont {Leswing}}, \bibinfo {author} {\bibfnamefont {E.~D.}\ \bibnamefont {Harder}},\ and\ \bibinfo {author} {\bibfnamefont {R.}~\bibnamefont {Abel}},\ }\bibfield  {title} {\enquote {\bibinfo {title} {Transferable neural network potential energy surfaces for closed-shell organic molecules: Extension to ions},}\ }\href {https://doi.org/10.1021/acs.jctc.1c00821} {\bibfield  {journal} {\bibinfo  {journal} {J. Chem. Theory Comput.}\ }\textbf {\bibinfo {volume} {18}},\ \bibinfo {pages} {2354--2366} (\bibinfo {year} {2022})}\BibitemShut {NoStop}%
\bibitem [{\citenamefont {Van~Voorhis}\ \emph {et~al.}(2010)\citenamefont {Van~Voorhis}, \citenamefont {Kowalczyk}, \citenamefont {Kaduk}, \citenamefont {Wang}, \citenamefont {Cheng},\ and\ \citenamefont {Wu}}]{VanVoorhis2010}%
  \BibitemOpen
  \bibfield  {author} {\bibinfo {author} {\bibfnamefont {T.}~\bibnamefont {Van~Voorhis}}, \bibinfo {author} {\bibfnamefont {T.}~\bibnamefont {Kowalczyk}}, \bibinfo {author} {\bibfnamefont {B.}~\bibnamefont {Kaduk}}, \bibinfo {author} {\bibfnamefont {L.-P.}\ \bibnamefont {Wang}}, \bibinfo {author} {\bibfnamefont {C.-L.}\ \bibnamefont {Cheng}},\ and\ \bibinfo {author} {\bibfnamefont {Q.}~\bibnamefont {Wu}},\ }\bibfield  {title} {\enquote {\bibinfo {title} {The diabatic picture of electron transfer, reaction barriers, and molecular dynamics},}\ }\href {https://doi.org/10.1146/annurev.physchem.012809.103324} {\bibfield  {journal} {\bibinfo  {journal} {Annu. Rev. Phys. Chem.}\ }\textbf {\bibinfo {volume} {61}},\ \bibinfo {pages} {149--170} (\bibinfo {year} {2010})}\BibitemShut {NoStop}%
\bibitem [{\citenamefont {Perdew}\ \emph {et~al.}(1982)\citenamefont {Perdew}, \citenamefont {Parr}, \citenamefont {Levy},\ and\ \citenamefont {Balduz}}]{Perdew1982}%
  \BibitemOpen
  \bibfield  {author} {\bibinfo {author} {\bibfnamefont {J.~P.}\ \bibnamefont {Perdew}}, \bibinfo {author} {\bibfnamefont {R.~G.}\ \bibnamefont {Parr}}, \bibinfo {author} {\bibfnamefont {M.}~\bibnamefont {Levy}},\ and\ \bibinfo {author} {\bibfnamefont {J.~L.}\ \bibnamefont {Balduz}},\ }\bibfield  {title} {\enquote {\bibinfo {title} {Density-functional theory for fractional particle number: Derivative discontinuities of the energy},}\ }\href {https://doi.org/10.1103/physrevlett.49.1691} {\bibfield  {journal} {\bibinfo  {journal} {Phys. Rev. Lett.}\ }\textbf {\bibinfo {volume} {49}},\ \bibinfo {pages} {1691--1694} (\bibinfo {year} {1982})}\BibitemShut {NoStop}%
\bibitem [{\citenamefont {Mori-Sánchez}\ and\ \citenamefont {Cohen}(2014)}]{MoriSanchez2014}%
  \BibitemOpen
  \bibfield  {author} {\bibinfo {author} {\bibfnamefont {P.}~\bibnamefont {Mori-Sánchez}}\ and\ \bibinfo {author} {\bibfnamefont {A.~J.}\ \bibnamefont {Cohen}},\ }\bibfield  {title} {\enquote {\bibinfo {title} {The derivative discontinuity of the exchange–correlation functional},}\ }\href {https://doi.org/10.1039/c4cp01170h} {\bibfield  {journal} {\bibinfo  {journal} {Phys. Chem. Chem. Phys.}\ }\textbf {\bibinfo {volume} {16}},\ \bibinfo {pages} {14378--14387} (\bibinfo {year} {2014})}\BibitemShut {NoStop}%
\bibitem [{\citenamefont {Sun}\ \emph {et~al.}(2017)\citenamefont {Sun}, \citenamefont {Berkelbach}, \citenamefont {Blunt}, \citenamefont {Booth}, \citenamefont {Guo}, \citenamefont {Li}, \citenamefont {Liu}, \citenamefont {McClain}, \citenamefont {Sayfutyarova}, \citenamefont {Sharma}, \citenamefont {Wouters},\ and\ \citenamefont {Chan}}]{Sun2017}%
  \BibitemOpen
  \bibfield  {author} {\bibinfo {author} {\bibfnamefont {Q.}~\bibnamefont {Sun}}, \bibinfo {author} {\bibfnamefont {T.~C.}\ \bibnamefont {Berkelbach}}, \bibinfo {author} {\bibfnamefont {N.~S.}\ \bibnamefont {Blunt}}, \bibinfo {author} {\bibfnamefont {G.~H.}\ \bibnamefont {Booth}}, \bibinfo {author} {\bibfnamefont {S.}~\bibnamefont {Guo}}, \bibinfo {author} {\bibfnamefont {Z.}~\bibnamefont {Li}}, \bibinfo {author} {\bibfnamefont {J.}~\bibnamefont {Liu}}, \bibinfo {author} {\bibfnamefont {J.~D.}\ \bibnamefont {McClain}}, \bibinfo {author} {\bibfnamefont {E.~R.}\ \bibnamefont {Sayfutyarova}}, \bibinfo {author} {\bibfnamefont {S.}~\bibnamefont {Sharma}}, \bibinfo {author} {\bibfnamefont {S.}~\bibnamefont {Wouters}},\ and\ \bibinfo {author} {\bibfnamefont {G.~K.}\ \bibnamefont {Chan}},\ }\bibfield  {title} {\enquote {\bibinfo {title} {{PySCF}: the {Python}‐based simulations of chemistry framework},}\ }\href {https://doi.org/10.1002/wcms.1340} {\bibfield  {journal} {\bibinfo  {journal} {WIREs Comput. Mol. Sci.}\
  }\textbf {\bibinfo {volume} {8}},\ \bibinfo {pages} {e1340} (\bibinfo {year} {2017})}\BibitemShut {NoStop}%
\bibitem [{\citenamefont {Sun}\ and\ \citenamefont {Chan}(2014)}]{Sun2014}%
  \BibitemOpen
  \bibfield  {author} {\bibinfo {author} {\bibfnamefont {Q.}~\bibnamefont {Sun}}\ and\ \bibinfo {author} {\bibfnamefont {G.~K.-L.}\ \bibnamefont {Chan}},\ }\bibfield  {title} {\enquote {\bibinfo {title} {Exact and optimal quantum mechanics/molecular mechanics boundaries},}\ }\href {https://doi.org/10.1021/ct500512f} {\bibfield  {journal} {\bibinfo  {journal} {Journal of Chemical Theory and Computation}\ }\textbf {\bibinfo {volume} {10}},\ \bibinfo {pages} {3784--3790} (\bibinfo {year} {2014})}\BibitemShut {NoStop}%
\bibitem [{\citenamefont {Becke}(1988)}]{Becke1988}%
  \BibitemOpen
  \bibfield  {author} {\bibinfo {author} {\bibfnamefont {A.~D.}\ \bibnamefont {Becke}},\ }\bibfield  {title} {\enquote {\bibinfo {title} {Density-functional exchange-energy approximation with correct asymptotic behavior},}\ }\href {https://doi.org/10.1103/physreva.38.3098} {\bibfield  {journal} {\bibinfo  {journal} {Phys. Rev. A}\ }\textbf {\bibinfo {volume} {38}},\ \bibinfo {pages} {3098--3100} (\bibinfo {year} {1988})}\BibitemShut {NoStop}%
\bibitem [{\citenamefont {Lee}, \citenamefont {Yang},\ and\ \citenamefont {Parr}(1988)}]{Lee1988}%
  \BibitemOpen
  \bibfield  {author} {\bibinfo {author} {\bibfnamefont {C.}~\bibnamefont {Lee}}, \bibinfo {author} {\bibfnamefont {W.}~\bibnamefont {Yang}},\ and\ \bibinfo {author} {\bibfnamefont {R.~G.}\ \bibnamefont {Parr}},\ }\bibfield  {title} {\enquote {\bibinfo {title} {Development of the {Colle-Salvetti} correlation-energy formula into a functional of the electron density},}\ }\href {https://doi.org/10.1103/physrevb.37.785} {\bibfield  {journal} {\bibinfo  {journal} {Phys. Rev. B}\ }\textbf {\bibinfo {volume} {37}},\ \bibinfo {pages} {785--789} (\bibinfo {year} {1988})}\BibitemShut {NoStop}%
\bibitem [{\citenamefont {Grimme}\ \emph {et~al.}(2010)\citenamefont {Grimme}, \citenamefont {Antony}, \citenamefont {Ehrlich},\ and\ \citenamefont {Krieg}}]{Grimme2010}%
  \BibitemOpen
  \bibfield  {author} {\bibinfo {author} {\bibfnamefont {S.}~\bibnamefont {Grimme}}, \bibinfo {author} {\bibfnamefont {J.}~\bibnamefont {Antony}}, \bibinfo {author} {\bibfnamefont {S.}~\bibnamefont {Ehrlich}},\ and\ \bibinfo {author} {\bibfnamefont {H.}~\bibnamefont {Krieg}},\ }\bibfield  {title} {\enquote {\bibinfo {title} {A consistent and accurate ab initio parametrization of density functional dispersion correction ({DFT-D}) for the 94 elements {H-Pu}},}\ }\href {https://doi.org/10.1063/1.3382344} {\bibfield  {journal} {\bibinfo  {journal} {J. Chem. Phys.}\ }\textbf {\bibinfo {volume} {132}},\ \bibinfo {pages} {154104} (\bibinfo {year} {2010})}\BibitemShut {NoStop}%
\bibitem [{\citenamefont {Grimme}, \citenamefont {Ehrlich},\ and\ \citenamefont {Goerigk}(2011)}]{Grimme2011}%
  \BibitemOpen
  \bibfield  {author} {\bibinfo {author} {\bibfnamefont {S.}~\bibnamefont {Grimme}}, \bibinfo {author} {\bibfnamefont {S.}~\bibnamefont {Ehrlich}},\ and\ \bibinfo {author} {\bibfnamefont {L.}~\bibnamefont {Goerigk}},\ }\bibfield  {title} {\enquote {\bibinfo {title} {Effect of the damping function in dispersion corrected density functional theory},}\ }\href {https://doi.org/10.1002/jcc.21759} {\bibfield  {journal} {\bibinfo  {journal} {J. Comput. Chem.}\ }\textbf {\bibinfo {volume} {32}},\ \bibinfo {pages} {1456--1465} (\bibinfo {year} {2011})}\BibitemShut {NoStop}%
\bibitem [{\citenamefont {Schäfer}, \citenamefont {Huber},\ and\ \citenamefont {Ahlrichs}(1994)}]{Schaefer1994}%
  \BibitemOpen
  \bibfield  {author} {\bibinfo {author} {\bibfnamefont {A.}~\bibnamefont {Schäfer}}, \bibinfo {author} {\bibfnamefont {C.}~\bibnamefont {Huber}},\ and\ \bibinfo {author} {\bibfnamefont {R.}~\bibnamefont {Ahlrichs}},\ }\bibfield  {title} {\enquote {\bibinfo {title} {Fully optimized contracted {Gaussian} basis sets of triple zeta valence quality for atoms {Li} to {Kr}},}\ }\href {https://doi.org/10.1063/1.467146} {\bibfield  {journal} {\bibinfo  {journal} {J. Chem. Phys.}\ }\textbf {\bibinfo {volume} {100}},\ \bibinfo {pages} {5829--5835} (\bibinfo {year} {1994})}\BibitemShut {NoStop}%
\bibitem [{\citenamefont {Weigend}\ and\ \citenamefont {Ahlrichs}(2005)}]{Weigend2005}%
  \BibitemOpen
  \bibfield  {author} {\bibinfo {author} {\bibfnamefont {F.}~\bibnamefont {Weigend}}\ and\ \bibinfo {author} {\bibfnamefont {R.}~\bibnamefont {Ahlrichs}},\ }\bibfield  {title} {\enquote {\bibinfo {title} {Balanced basis sets of split valence, triple zeta valence and quadruple zeta valence quality for {H} to {Rn}: Design and assessment of accuracy},}\ }\href {https://doi.org/10.1039/b508541a} {\bibfield  {journal} {\bibinfo  {journal} {Phys. Chem. Chem. Phys.}\ }\textbf {\bibinfo {volume} {7}},\ \bibinfo {pages} {3297} (\bibinfo {year} {2005})}\BibitemShut {NoStop}%
\bibitem [{\citenamefont {Siddique}\ \emph {et~al.}(2020)\citenamefont {Siddique}, \citenamefont {Barbatti}, \citenamefont {Cui}, \citenamefont {Lischka},\ and\ \citenamefont {Aquino}}]{Siddique2020}%
  \BibitemOpen
  \bibfield  {author} {\bibinfo {author} {\bibfnamefont {F.}~\bibnamefont {Siddique}}, \bibinfo {author} {\bibfnamefont {M.}~\bibnamefont {Barbatti}}, \bibinfo {author} {\bibfnamefont {Z.}~\bibnamefont {Cui}}, \bibinfo {author} {\bibfnamefont {H.}~\bibnamefont {Lischka}},\ and\ \bibinfo {author} {\bibfnamefont {A.~J.~A.}\ \bibnamefont {Aquino}},\ }\bibfield  {title} {\enquote {\bibinfo {title} {Nonadiabatic dynamics of charge-transfer states using the anthracene–tetracyanoethylene complex as a prototype},}\ }\href {https://doi.org/10.1021/acs.jpca.0c01900} {\bibfield  {journal} {\bibinfo  {journal} {J. Phys. Chem. A}\ }\textbf {\bibinfo {volume} {124}},\ \bibinfo {pages} {3347--3357} (\bibinfo {year} {2020})}\BibitemShut {NoStop}%
\bibitem [{\citenamefont {Ebadi}, \citenamefont {Brandell},\ and\ \citenamefont {Araujo}(2016)}]{Ebadi2016}%
  \BibitemOpen
  \bibfield  {author} {\bibinfo {author} {\bibfnamefont {M.}~\bibnamefont {Ebadi}}, \bibinfo {author} {\bibfnamefont {D.}~\bibnamefont {Brandell}},\ and\ \bibinfo {author} {\bibfnamefont {C.~M.}\ \bibnamefont {Araujo}},\ }\bibfield  {title} {\enquote {\bibinfo {title} {Electrolyte decomposition on {Li}-metal surfaces from first-principles theory},}\ }\href {https://doi.org/10.1063/1.4967810} {\bibfield  {journal} {\bibinfo  {journal} {J. Chem. Phys.}\ }\textbf {\bibinfo {volume} {145}},\ \bibinfo {pages} {204701} (\bibinfo {year} {2016})}\BibitemShut {NoStop}%
\bibitem [{\citenamefont {Brennan}\ \emph {et~al.}(2017)\citenamefont {Brennan}, \citenamefont {Breedon}, \citenamefont {Best}, \citenamefont {Morishita},\ and\ \citenamefont {Spencer}}]{Brennan2017}%
  \BibitemOpen
  \bibfield  {author} {\bibinfo {author} {\bibfnamefont {M.~D.}\ \bibnamefont {Brennan}}, \bibinfo {author} {\bibfnamefont {M.}~\bibnamefont {Breedon}}, \bibinfo {author} {\bibfnamefont {A.~S.}\ \bibnamefont {Best}}, \bibinfo {author} {\bibfnamefont {T.}~\bibnamefont {Morishita}},\ and\ \bibinfo {author} {\bibfnamefont {M.~J.}\ \bibnamefont {Spencer}},\ }\bibfield  {title} {\enquote {\bibinfo {title} {Surface reactions of ethylene carbonate and propylene carbonate on the {Li(001)} surface},}\ }\href {https://doi.org/10.1016/j.electacta.2017.04.163} {\bibfield  {journal} {\bibinfo  {journal} {Electrochim. Acta}\ }\textbf {\bibinfo {volume} {243}},\ \bibinfo {pages} {320--330} (\bibinfo {year} {2017})}\BibitemShut {NoStop}%
\bibitem [{\citenamefont {Camacho-Forero}\ \emph {et~al.}(2015)\citenamefont {Camacho-Forero}, \citenamefont {Smith}, \citenamefont {Bertolini},\ and\ \citenamefont {Balbuena}}]{CamachoForero2015}%
  \BibitemOpen
  \bibfield  {author} {\bibinfo {author} {\bibfnamefont {L.~E.}\ \bibnamefont {Camacho-Forero}}, \bibinfo {author} {\bibfnamefont {T.~W.}\ \bibnamefont {Smith}}, \bibinfo {author} {\bibfnamefont {S.}~\bibnamefont {Bertolini}},\ and\ \bibinfo {author} {\bibfnamefont {P.~B.}\ \bibnamefont {Balbuena}},\ }\bibfield  {title} {\enquote {\bibinfo {title} {Reactivity at the lithium–metal anode surface of lithium–sulfur batteries},}\ }\href {https://doi.org/10.1021/acs.jpcc.5b08254} {\bibfield  {journal} {\bibinfo  {journal} {J. Phys. Chem. C}\ }\textbf {\bibinfo {volume} {119}},\ \bibinfo {pages} {26828--26839} (\bibinfo {year} {2015})}\BibitemShut {NoStop}%
\bibitem [{\citenamefont {Giannozzi}\ \emph {et~al.}(2009)\citenamefont {Giannozzi}, \citenamefont {Baroni}, \citenamefont {Bonini}, \citenamefont {Calandra}, \citenamefont {Car}, \citenamefont {Cavazzoni}, \citenamefont {Ceresoli}, \citenamefont {Chiarotti}, \citenamefont {Cococcioni}, \citenamefont {Dabo}, \citenamefont {Dal~Corso}, \citenamefont {de~Gironcoli}, \citenamefont {Fabris}, \citenamefont {Fratesi}, \citenamefont {Gebauer}, \citenamefont {Gerstmann}, \citenamefont {Gougoussis}, \citenamefont {Kokalj}, \citenamefont {Lazzeri}, \citenamefont {Martin-Samos}, \citenamefont {Marzari}, \citenamefont {Mauri}, \citenamefont {Mazzarello}, \citenamefont {Paolini}, \citenamefont {Pasquarello}, \citenamefont {Paulatto}, \citenamefont {Sbraccia}, \citenamefont {Scandolo}, \citenamefont {Sclauzero}, \citenamefont {Seitsonen}, \citenamefont {Smogunov}, \citenamefont {Umari},\ and\ \citenamefont {Wentzcovitch}}]{Giannozzi2009}%
  \BibitemOpen
  \bibfield  {author} {\bibinfo {author} {\bibfnamefont {P.}~\bibnamefont {Giannozzi}}, \bibinfo {author} {\bibfnamefont {S.}~\bibnamefont {Baroni}}, \bibinfo {author} {\bibfnamefont {N.}~\bibnamefont {Bonini}}, \bibinfo {author} {\bibfnamefont {M.}~\bibnamefont {Calandra}}, \bibinfo {author} {\bibfnamefont {R.}~\bibnamefont {Car}}, \bibinfo {author} {\bibfnamefont {C.}~\bibnamefont {Cavazzoni}}, \bibinfo {author} {\bibfnamefont {D.}~\bibnamefont {Ceresoli}}, \bibinfo {author} {\bibfnamefont {G.~L.}\ \bibnamefont {Chiarotti}}, \bibinfo {author} {\bibfnamefont {M.}~\bibnamefont {Cococcioni}}, \bibinfo {author} {\bibfnamefont {I.}~\bibnamefont {Dabo}}, \bibinfo {author} {\bibfnamefont {A.}~\bibnamefont {Dal~Corso}}, \bibinfo {author} {\bibfnamefont {S.}~\bibnamefont {de~Gironcoli}}, \bibinfo {author} {\bibfnamefont {S.}~\bibnamefont {Fabris}}, \bibinfo {author} {\bibfnamefont {G.}~\bibnamefont {Fratesi}}, \bibinfo {author} {\bibfnamefont {R.}~\bibnamefont {Gebauer}}, \bibinfo {author} {\bibfnamefont
  {U.}~\bibnamefont {Gerstmann}}, \bibinfo {author} {\bibfnamefont {C.}~\bibnamefont {Gougoussis}}, \bibinfo {author} {\bibfnamefont {A.}~\bibnamefont {Kokalj}}, \bibinfo {author} {\bibfnamefont {M.}~\bibnamefont {Lazzeri}}, \bibinfo {author} {\bibfnamefont {L.}~\bibnamefont {Martin-Samos}}, \bibinfo {author} {\bibfnamefont {N.}~\bibnamefont {Marzari}}, \bibinfo {author} {\bibfnamefont {F.}~\bibnamefont {Mauri}}, \bibinfo {author} {\bibfnamefont {R.}~\bibnamefont {Mazzarello}}, \bibinfo {author} {\bibfnamefont {S.}~\bibnamefont {Paolini}}, \bibinfo {author} {\bibfnamefont {A.}~\bibnamefont {Pasquarello}}, \bibinfo {author} {\bibfnamefont {L.}~\bibnamefont {Paulatto}}, \bibinfo {author} {\bibfnamefont {C.}~\bibnamefont {Sbraccia}}, \bibinfo {author} {\bibfnamefont {S.}~\bibnamefont {Scandolo}}, \bibinfo {author} {\bibfnamefont {G.}~\bibnamefont {Sclauzero}}, \bibinfo {author} {\bibfnamefont {A.~P.}\ \bibnamefont {Seitsonen}}, \bibinfo {author} {\bibfnamefont {A.}~\bibnamefont {Smogunov}}, \bibinfo {author}
  {\bibfnamefont {P.}~\bibnamefont {Umari}},\ and\ \bibinfo {author} {\bibfnamefont {R.~M.}\ \bibnamefont {Wentzcovitch}},\ }\bibfield  {title} {\enquote {\bibinfo {title} {{QUANTUM ESPRESSO}: a modular and open-source software project for quantum simulations of materials},}\ }\href {https://doi.org/10.1088/0953-8984/21/39/395502} {\bibfield  {journal} {\bibinfo  {journal} {J. Phys. Condens. Matter}\ }\textbf {\bibinfo {volume} {21}},\ \bibinfo {pages} {395502} (\bibinfo {year} {2009})}\BibitemShut {NoStop}%
\bibitem [{\citenamefont {Giannozzi}\ \emph {et~al.}(2017)\citenamefont {Giannozzi}, \citenamefont {Andreussi}, \citenamefont {Brumme}, \citenamefont {Bunau}, \citenamefont {Buongiorno~Nardelli}, \citenamefont {Calandra}, \citenamefont {Car}, \citenamefont {Cavazzoni}, \citenamefont {Ceresoli}, \citenamefont {Cococcioni}, \citenamefont {Colonna}, \citenamefont {Carnimeo}, \citenamefont {Dal~Corso}, \citenamefont {de~Gironcoli}, \citenamefont {Delugas}, \citenamefont {DiStasio}, \citenamefont {Ferretti}, \citenamefont {Floris}, \citenamefont {Fratesi}, \citenamefont {Fugallo}, \citenamefont {Gebauer}, \citenamefont {Gerstmann}, \citenamefont {Giustino}, \citenamefont {Gorni}, \citenamefont {Jia}, \citenamefont {Kawamura}, \citenamefont {Ko}, \citenamefont {Kokalj}, \citenamefont {Küçükbenli}, \citenamefont {Lazzeri}, \citenamefont {Marsili}, \citenamefont {Marzari}, \citenamefont {Mauri}, \citenamefont {Nguyen}, \citenamefont {Nguyen}, \citenamefont {Otero-de-la Roza}, \citenamefont {Paulatto},
  \citenamefont {Poncé}, \citenamefont {Rocca}, \citenamefont {Sabatini}, \citenamefont {Santra}, \citenamefont {Schlipf}, \citenamefont {Seitsonen}, \citenamefont {Smogunov}, \citenamefont {Timrov}, \citenamefont {Thonhauser}, \citenamefont {Umari}, \citenamefont {Vast}, \citenamefont {Wu},\ and\ \citenamefont {Baroni}}]{Giannozzi2017}%
  \BibitemOpen
  \bibfield  {author} {\bibinfo {author} {\bibfnamefont {P.}~\bibnamefont {Giannozzi}}, \bibinfo {author} {\bibfnamefont {O.}~\bibnamefont {Andreussi}}, \bibinfo {author} {\bibfnamefont {T.}~\bibnamefont {Brumme}}, \bibinfo {author} {\bibfnamefont {O.}~\bibnamefont {Bunau}}, \bibinfo {author} {\bibfnamefont {M.}~\bibnamefont {Buongiorno~Nardelli}}, \bibinfo {author} {\bibfnamefont {M.}~\bibnamefont {Calandra}}, \bibinfo {author} {\bibfnamefont {R.}~\bibnamefont {Car}}, \bibinfo {author} {\bibfnamefont {C.}~\bibnamefont {Cavazzoni}}, \bibinfo {author} {\bibfnamefont {D.}~\bibnamefont {Ceresoli}}, \bibinfo {author} {\bibfnamefont {M.}~\bibnamefont {Cococcioni}}, \bibinfo {author} {\bibfnamefont {N.}~\bibnamefont {Colonna}}, \bibinfo {author} {\bibfnamefont {I.}~\bibnamefont {Carnimeo}}, \bibinfo {author} {\bibfnamefont {A.}~\bibnamefont {Dal~Corso}}, \bibinfo {author} {\bibfnamefont {S.}~\bibnamefont {de~Gironcoli}}, \bibinfo {author} {\bibfnamefont {P.}~\bibnamefont {Delugas}}, \bibinfo {author} {\bibfnamefont
  {R.~A.}\ \bibnamefont {DiStasio}}, \bibinfo {author} {\bibfnamefont {A.}~\bibnamefont {Ferretti}}, \bibinfo {author} {\bibfnamefont {A.}~\bibnamefont {Floris}}, \bibinfo {author} {\bibfnamefont {G.}~\bibnamefont {Fratesi}}, \bibinfo {author} {\bibfnamefont {G.}~\bibnamefont {Fugallo}}, \bibinfo {author} {\bibfnamefont {R.}~\bibnamefont {Gebauer}}, \bibinfo {author} {\bibfnamefont {U.}~\bibnamefont {Gerstmann}}, \bibinfo {author} {\bibfnamefont {F.}~\bibnamefont {Giustino}}, \bibinfo {author} {\bibfnamefont {T.}~\bibnamefont {Gorni}}, \bibinfo {author} {\bibfnamefont {J.}~\bibnamefont {Jia}}, \bibinfo {author} {\bibfnamefont {M.}~\bibnamefont {Kawamura}}, \bibinfo {author} {\bibfnamefont {H.-Y.}\ \bibnamefont {Ko}}, \bibinfo {author} {\bibfnamefont {A.}~\bibnamefont {Kokalj}}, \bibinfo {author} {\bibfnamefont {E.}~\bibnamefont {Küçükbenli}}, \bibinfo {author} {\bibfnamefont {M.}~\bibnamefont {Lazzeri}}, \bibinfo {author} {\bibfnamefont {M.}~\bibnamefont {Marsili}}, \bibinfo {author} {\bibfnamefont
  {N.}~\bibnamefont {Marzari}}, \bibinfo {author} {\bibfnamefont {F.}~\bibnamefont {Mauri}}, \bibinfo {author} {\bibfnamefont {N.~L.}\ \bibnamefont {Nguyen}}, \bibinfo {author} {\bibfnamefont {H.-V.}\ \bibnamefont {Nguyen}}, \bibinfo {author} {\bibfnamefont {A.}~\bibnamefont {Otero-de-la Roza}}, \bibinfo {author} {\bibfnamefont {L.}~\bibnamefont {Paulatto}}, \bibinfo {author} {\bibfnamefont {S.}~\bibnamefont {Poncé}}, \bibinfo {author} {\bibfnamefont {D.}~\bibnamefont {Rocca}}, \bibinfo {author} {\bibfnamefont {R.}~\bibnamefont {Sabatini}}, \bibinfo {author} {\bibfnamefont {B.}~\bibnamefont {Santra}}, \bibinfo {author} {\bibfnamefont {M.}~\bibnamefont {Schlipf}}, \bibinfo {author} {\bibfnamefont {A.~P.}\ \bibnamefont {Seitsonen}}, \bibinfo {author} {\bibfnamefont {A.}~\bibnamefont {Smogunov}}, \bibinfo {author} {\bibfnamefont {I.}~\bibnamefont {Timrov}}, \bibinfo {author} {\bibfnamefont {T.}~\bibnamefont {Thonhauser}}, \bibinfo {author} {\bibfnamefont {P.}~\bibnamefont {Umari}}, \bibinfo {author}
  {\bibfnamefont {N.}~\bibnamefont {Vast}}, \bibinfo {author} {\bibfnamefont {X.}~\bibnamefont {Wu}},\ and\ \bibinfo {author} {\bibfnamefont {S.}~\bibnamefont {Baroni}},\ }\bibfield  {title} {\enquote {\bibinfo {title} {Advanced capabilities for materials modelling with {Quantum} {ESPRESSO}},}\ }\href {https://doi.org/10.1088/1361-648x/aa8f79} {\bibfield  {journal} {\bibinfo  {journal} {J. Phys. Condens. Matter}\ }\textbf {\bibinfo {volume} {29}},\ \bibinfo {pages} {465901} (\bibinfo {year} {2017})}\BibitemShut {NoStop}%
\bibitem [{\citenamefont {Perdew}, \citenamefont {Burke},\ and\ \citenamefont {Ernzerhof}(1996)}]{Perdew1996}%
  \BibitemOpen
  \bibfield  {author} {\bibinfo {author} {\bibfnamefont {J.~P.}\ \bibnamefont {Perdew}}, \bibinfo {author} {\bibfnamefont {K.}~\bibnamefont {Burke}},\ and\ \bibinfo {author} {\bibfnamefont {M.}~\bibnamefont {Ernzerhof}},\ }\bibfield  {title} {\enquote {\bibinfo {title} {Generalized gradient approximation made simple},}\ }\href {https://doi.org/10.1103/physrevlett.77.3865} {\bibfield  {journal} {\bibinfo  {journal} {Phys. Rev. Lett.}\ }\textbf {\bibinfo {volume} {77}},\ \bibinfo {pages} {3865--3868} (\bibinfo {year} {1996})}\BibitemShut {NoStop}%
\bibitem [{\citenamefont {Perdew}, \citenamefont {Burke},\ and\ \citenamefont {Ernzerhof}(1997)}]{Perdew1997}%
  \BibitemOpen
  \bibfield  {author} {\bibinfo {author} {\bibfnamefont {J.~P.}\ \bibnamefont {Perdew}}, \bibinfo {author} {\bibfnamefont {K.}~\bibnamefont {Burke}},\ and\ \bibinfo {author} {\bibfnamefont {M.}~\bibnamefont {Ernzerhof}},\ }\bibfield  {title} {\enquote {\bibinfo {title} {Generalized gradient approximation made simple [phys. rev. lett. 77, 3865 (1996)]},}\ }\href {https://doi.org/10.1103/physrevlett.78.1396} {\bibfield  {journal} {\bibinfo  {journal} {Phys. Rev. Lett.}\ }\textbf {\bibinfo {volume} {78}},\ \bibinfo {pages} {1396--1396} (\bibinfo {year} {1997})}\BibitemShut {NoStop}%
\bibitem [{\citenamefont {Ye}\ and\ \citenamefont {Berkelbach}(2022)}]{Ye2022}%
  \BibitemOpen
  \bibfield  {author} {\bibinfo {author} {\bibfnamefont {H.-Z.}\ \bibnamefont {Ye}}\ and\ \bibinfo {author} {\bibfnamefont {T.~C.}\ \bibnamefont {Berkelbach}},\ }\bibfield  {title} {\enquote {\bibinfo {title} {Correlation-consistent gaussian basis sets for solids made simple},}\ }\href {https://doi.org/10.1021/acs.jctc.1c01245} {\bibfield  {journal} {\bibinfo  {journal} {J. Chem. Theory Comput.}\ }\textbf {\bibinfo {volume} {18}},\ \bibinfo {pages} {1595--1606} (\bibinfo {year} {2022})}\BibitemShut {NoStop}%
\bibitem [{\citenamefont {Baeck}\ and\ \citenamefont {An}(2017)}]{Baeck2017}%
  \BibitemOpen
  \bibfield  {author} {\bibinfo {author} {\bibfnamefont {K.~K.}\ \bibnamefont {Baeck}}\ and\ \bibinfo {author} {\bibfnamefont {H.}~\bibnamefont {An}},\ }\bibfield  {title} {\enquote {\bibinfo {title} {Practical approximation of the non-adiabatic coupling terms for same-symmetry interstate crossings by using adiabatic potential energies only},}\ }\href {https://doi.org/10.1063/1.4975323} {\bibfield  {journal} {\bibinfo  {journal} {J. Chem. Phys.}\ }\textbf {\bibinfo {volume} {146}} (\bibinfo {year} {2017}),\ 10.1063/1.4975323}\BibitemShut {NoStop}%
\bibitem [{\citenamefont {Shu}\ \emph {et~al.}(2022)\citenamefont {Shu}, \citenamefont {Zhang}, \citenamefont {Chen}, \citenamefont {Sun}, \citenamefont {Huang},\ and\ \citenamefont {Truhlar}}]{Shu2022}%
  \BibitemOpen
  \bibfield  {author} {\bibinfo {author} {\bibfnamefont {Y.}~\bibnamefont {Shu}}, \bibinfo {author} {\bibfnamefont {L.}~\bibnamefont {Zhang}}, \bibinfo {author} {\bibfnamefont {X.}~\bibnamefont {Chen}}, \bibinfo {author} {\bibfnamefont {S.}~\bibnamefont {Sun}}, \bibinfo {author} {\bibfnamefont {Y.}~\bibnamefont {Huang}},\ and\ \bibinfo {author} {\bibfnamefont {D.~G.}\ \bibnamefont {Truhlar}},\ }\bibfield  {title} {\enquote {\bibinfo {title} {Nonadiabatic dynamics algorithms with only potential energies and gradients: Curvature-driven coherent switching with decay of mixing and curvature-driven trajectory surface hopping},}\ }\href {https://doi.org/10.1021/acs.jctc.1c01080} {\bibfield  {journal} {\bibinfo  {journal} {J. Chem. Theory Comput.}\ }\textbf {\bibinfo {volume} {18}},\ \bibinfo {pages} {1320--1328} (\bibinfo {year} {2022})}\BibitemShut {NoStop}%
\bibitem [{\citenamefont {Vuilleumier}\ and\ \citenamefont {Borgis}(1998)}]{Vuilleumier1998}%
  \BibitemOpen
  \bibfield  {author} {\bibinfo {author} {\bibfnamefont {R.}~\bibnamefont {Vuilleumier}}\ and\ \bibinfo {author} {\bibfnamefont {D.}~\bibnamefont {Borgis}},\ }\bibfield  {title} {\enquote {\bibinfo {title} {An extended empirical valence bond model for describing proton transfer in h+(h2o)n clusters and liquid water},}\ }\href {https://doi.org/10.1016/s0009-2614(97)01365-1} {\bibfield  {journal} {\bibinfo  {journal} {Chem. Phys. Lett.}\ }\textbf {\bibinfo {volume} {284}},\ \bibinfo {pages} {71--77} (\bibinfo {year} {1998})}\BibitemShut {NoStop}%
\bibitem [{\citenamefont {Schmitt}\ and\ \citenamefont {Voth}(1999)}]{Schmitt1999}%
  \BibitemOpen
  \bibfield  {author} {\bibinfo {author} {\bibfnamefont {U.~W.}\ \bibnamefont {Schmitt}}\ and\ \bibinfo {author} {\bibfnamefont {G.~A.}\ \bibnamefont {Voth}},\ }\bibfield  {title} {\enquote {\bibinfo {title} {The computer simulation of proton transport in water},}\ }\href {https://doi.org/10.1063/1.480032} {\bibfield  {journal} {\bibinfo  {journal} {J. Chem. Phys.}\ }\textbf {\bibinfo {volume} {111}},\ \bibinfo {pages} {9361--9381} (\bibinfo {year} {1999})}\BibitemShut {NoStop}%
\bibitem [{\citenamefont {Day}\ \emph {et~al.}(2002)\citenamefont {Day}, \citenamefont {Soudackov}, \citenamefont {Čuma}, \citenamefont {Schmitt},\ and\ \citenamefont {Voth}}]{Day2002}%
  \BibitemOpen
  \bibfield  {author} {\bibinfo {author} {\bibfnamefont {T.~J.~F.}\ \bibnamefont {Day}}, \bibinfo {author} {\bibfnamefont {A.~V.}\ \bibnamefont {Soudackov}}, \bibinfo {author} {\bibfnamefont {M.}~\bibnamefont {Čuma}}, \bibinfo {author} {\bibfnamefont {U.~W.}\ \bibnamefont {Schmitt}},\ and\ \bibinfo {author} {\bibfnamefont {G.~A.}\ \bibnamefont {Voth}},\ }\bibfield  {title} {\enquote {\bibinfo {title} {A second generation multistate empirical valence bond model for proton transport in aqueous systems},}\ }\href {https://doi.org/10.1063/1.1497157} {\bibfield  {journal} {\bibinfo  {journal} {J. Chem. Phys.}\ }\textbf {\bibinfo {volume} {117}},\ \bibinfo {pages} {5839--5849} (\bibinfo {year} {2002})}\BibitemShut {NoStop}%
\bibitem [{\citenamefont {Rick}, \citenamefont {Stuart},\ and\ \citenamefont {Berne}(1994)}]{Rick1994}%
  \BibitemOpen
  \bibfield  {author} {\bibinfo {author} {\bibfnamefont {S.~W.}\ \bibnamefont {Rick}}, \bibinfo {author} {\bibfnamefont {S.~J.}\ \bibnamefont {Stuart}},\ and\ \bibinfo {author} {\bibfnamefont {B.~J.}\ \bibnamefont {Berne}},\ }\bibfield  {title} {\enquote {\bibinfo {title} {Dynamical fluctuating charge force fields: Application to liquid water},}\ }\href {https://doi.org/10.1063/1.468398} {\bibfield  {journal} {\bibinfo  {journal} {J. Chem. Phys.}\ }\textbf {\bibinfo {volume} {101}},\ \bibinfo {pages} {6141--6156} (\bibinfo {year} {1994})}\BibitemShut {NoStop}%
\bibitem [{\citenamefont {Shimizu}\ \emph {et~al.}(2004)\citenamefont {Shimizu}, \citenamefont {Chaimovich}, \citenamefont {Farah}, \citenamefont {Dias},\ and\ \citenamefont {Bostick}}]{Shimizu2004}%
  \BibitemOpen
  \bibfield  {author} {\bibinfo {author} {\bibfnamefont {K.}~\bibnamefont {Shimizu}}, \bibinfo {author} {\bibfnamefont {H.}~\bibnamefont {Chaimovich}}, \bibinfo {author} {\bibfnamefont {J.~P.~S.}\ \bibnamefont {Farah}}, \bibinfo {author} {\bibfnamefont {L.~G.}\ \bibnamefont {Dias}},\ and\ \bibinfo {author} {\bibfnamefont {D.~L.}\ \bibnamefont {Bostick}},\ }\bibfield  {title} {\enquote {\bibinfo {title} {Calculation of the dipole moment for polypeptides using the generalized {Born}-electronegativity equalization method: Results in vacuum and continuum-dielectric solvent},}\ }\href {https://doi.org/10.1021/jp037315w} {\bibfield  {journal} {\bibinfo  {journal} {J. Phys. Chem. B}\ }\textbf {\bibinfo {volume} {108}},\ \bibinfo {pages} {4171--4177} (\bibinfo {year} {2004})}\BibitemShut {NoStop}%
\end{thebibliography}
\end{document}